\documentclass[a4paper,11pt]{article}
\pdfoutput=1

\usepackage{jheppub}
\usepackage{booktabs}
\usepackage{graphicx}
\usepackage{amsmath}
\usepackage{amssymb}
\usepackage{amsfonts}
\usepackage{dcolumn}
\usepackage[normalem]{ulem}
\usepackage{bm}
\usepackage[dvipsnames]{xcolor}
\usepackage[utf8]{inputenc}
\usepackage{subfigure}
\usepackage{gensymb}
\usepackage{cleveref}
\usepackage{colortbl}
\usepackage{tabularx}
\usepackage{tabulary}
\usepackage{tabularray}
\usepackage{tablefootnote}
\usepackage{enumitem}

\usepackage[T1]{fontenc}
\usepackage{pstricks}
\usepackage{color}
\usepackage{multirow}
\usepackage{slashed}
\usepackage{mathtools}
\usepackage{mathrsfs}
\usepackage{bbold}
\usepackage[force]{feynmp-auto}
\allowdisplaybreaks[4]

\usepackage{verbatim}

\DeclareMathAlphabet{\mathpzc}{OT1}{pzc}{m}{it}

\hypersetup{colorlinks,linkcolor={blue},citecolor={teal},urlcolor={violet}}

\newcommand{\SOTON}{School of Physics and Astronomy, University of Southampton, SO17 1BJ Southampton, United Kingdom}
\newcommand{\calA}{\mathcal{A}}
\newcommand{\calK}{\mathcal{K}}
\newcommand{\calW}{\mathcal{W}}

\newcommand{\imt}{{\rm Im}\,\tau}

\newcommand{\fm}{\mathfrak{m}}

\newcommand{\I}{{\rm i}}
\newcommand{\plk}{M_{\rm p}}
\newcommand{\varall}{\tau, \overline{\tau},S,\overline{S}}

\newcommand{\figref}[1]{Fig.~\ref{#1}}

\allowdisplaybreaks[4]

\begin{document}

\title{Modular domain walls and gravitational waves}

\author[a]{Stephen F. King\note{\url{https://orcid.org/0000-0002-4351-7507}},}

\author[a]{Xin Wang\note{\url{https://orcid.org/0000-0003-4292-460X}}}

\author{and}

\author[b]{Ye-Ling Zhou}\note{\url{https://orcid.org/0000-0002-3664-9472}}

\affiliation[a]{\SOTON}
\affiliation[b]{School of Fundamental Physics and Mathematical Sciences, Hangzhou Institute for Advanced Study, UCAS,
Hangzhou 310024, China}

\emailAdd{king@soton.ac.uk}
\emailAdd{xin.wang@soton.ac.uk}
\emailAdd{zhouyeling@ucas.ac.cn}

\abstract{
We discuss modular domain walls and gravitational waves in a class of supersymmetric models where quark and lepton flavour symmetry emerges from modular symmetry. In such models a single modulus field $\tau$ is often assumed to be stabilised at or near certain fixed point values such as 
$\tau = {\rm i}$ and $\tau = \omega$ (the cube root of unity), in its fundamental domain. We show that, in the global supersymmetry limit of certain classes of potentials, the vacua at these fixed points may be degenerate, leading to the formation of modular domain walls in the early Universe. Taking supergravity effects into account, in the background of a fixed dilaton field $S$, the degeneracy may be lifted, leading to a bias term in the potential allowing the domain walls to collapse. We study the resulting gravitational wave spectra arising from the dynamics of such modular domain walls, and assess their observability by current and future experiments, as a window into modular flavour symmetry.}

\maketitle

\section{Introduction} \label{sec:intro}

The curious pattern of quark and lepton (including neutrino) masses and mixing remains one of the major puzzles of the Standard Model (SM)~\cite{Xing:2020ijf}.
Many approaches to this so called flavour problem involve the introduction of a flavour symmetry which is spontaneously broken by new Higgs scalar fields called flavons, whose vacuum alignment is controlled by further scalar fields called driving fields (for a review see, e.g., Ref.~\cite{King:2013eh}).

It has been suggested that finite flavour symmetries might arise from an infinite modular symmetry necessarily broken by a single complex modulus field $\tau$, in a bottom-up approach~\cite{Feruglio:2017spp} using ideas borrowed from string theory~\cite{Lauer:1989ax,Ferrara:1989bc,Ferrara:1989qb}.
For instance, 
the infinite modular symmetry, ${\rm PSL}(2,\mathbb{Z})$, with  its series of infinite normal subgroups called the principle congruence subgroups $\Gamma (N)$ of level $N$, permits finite quotient groups $\Gamma_N \cong {\rm PSL}(2,\mathbb{Z})/\Gamma (N)$ identified with flavour groups. 
For low integer levels $N=3,4,5$ familiar flavour symmetries emerge,  $\Gamma_3 \cong A_4$~\cite{Feruglio:2017spp,Kobayashi:2018vbk, Criado:2018thu, Kobayashi:2018scp,deAnda:2018ecu,Okada:2019uoy, Ding:2019zxk,Zhang:2019ngf,Kobayashi:2019gtp,Wang:2019xbo,Okada:2020rjb,Yao:2020qyy,Chen:2021zty, Kobayashi:2021pav, Kang:2022psa, CentellesChulia:2023osj}, $\Gamma_4 \cong S_4$~\cite{Penedo:2018nmg,Novichkov:2018ovf,Kobayashi:2019mna,Wang:2019ovr,Zhang:2021olk}, 
$\Gamma_5 \cong A_5$~\cite{Novichkov:2018nkm,Ding:2019xna,Criado:2019tzk}. This approach has been widely developed in recent literature and many other related ideas and examples have been proposed (for reviews see, e.g., Refs.~\cite{Kobayashi:2023zzc, Ding:2023htn}).

In many bottom-up models, the only flavon present is the single modulus field $\tau$, whose
vacuum expectation value (VEV) fixes the value of Yukawa couplings which form representations of the finite modular flavour symmetry and
are modular forms, leading to predictive models~\cite{Feruglio:2017spp}. Without loss of generality, $\tau$ may be considered in its fundamental domain in the upper-half complex plane, within which 
there are three fixed points where a discrete subgroup of the modular symmetry is preserved.
The three fixed points are: $\tau = {\rm i} $ (preserving $Z_2^S$), $\tau = \omega = e^{2\pi{\rm i}/3}$ (preserving $Z_3^{ST}$), and 
$\tau = {\rm i} \infty $ (preserving $Z_N^T$), where $S,T$ generate the modular symmetry~\cite{Feruglio:2017spp}.
These fixed points seem to have phenomenological relevance in particular models
(see, e.g., Refs.~\cite{Novichkov:2018ovf,Novichkov:2018yse,Ding:2019gof, deMedeirosVarzielas:2020kji}).

From a theoretical point of view, moduli fields should be stabilised at the minima of some potentials, and there is extensive string theory literature concerned with this problem. For example, in heterotic string theory, gaugino condensation~\cite{Dine:1985rz, Nilles:1982ik, Ferrara:1982qs} combined with 
threshold corrections~\cite{Kaplunovsky:1987rp, Dixon:1990pc,  Antoniadis:1991fh, Antoniadis:1992rq,Antoniadis:1992sa,Kaplunovsky:1995jw} can lead to non-trivial vacua which may be used to stabilise moduli fields~\cite{Font:1990nt, Cvetic:1991qm, Cicoli:2013rwa, Gonzalo:2018guu}. More recently these issues have been revisited in the bottom-up approach
~\cite{Kobayashi:2019xvz,  Ishiguro:2020tmo, Novichkov:2022wvg,Funakoshi:2024yxg}. Many of these vacua are anti-de Sitter (AdS), but approaches to achieve de Sitter (dS) vacua have also been proposed~\cite{Ishiguro:2022pde,Leedom:2022zdm,Knapp-Perez:2023nty,King:2023snq}. For example, by including Shenker-like effects~\cite{Shenker:1990} as non-perturbative corrections to the dilaton K\"ahler potential, one may obtain metastable dS vacua at the fixed points $\tau = {\rm i}$ and $\omega$~\cite{Leedom:2022zdm,King:2023snq}. Interestingly, it has been found that such potentials may also lead to the slow-roll inflation~\cite{Ding:2024neh, King:2024ssx}.

In this paper, we show that, in the global supersymmetry limit of certain classes of potentials, the vacua at the fixed points $\tau = {\rm i}$ and $\tau =\omega$ may be degenerate, leading to the formation of modular domain walls (DWs) in the early Universe~\cite{Cvetic:1992bf,Cvetic:1992cv,Cvetic:1996vr,Cvetic:1991vp}. Traditionally, DWs arise from theories in which a discrete symmetry such as $Z_2$ is spontaneously broken~\cite{Zeldovich:1974uw,Kibble:1976sj,Vilenkin:1984ib}. However, modular DWs are very different from DWs from ordinary discrete symmetry breaking in field theory. For DWs in field theory, the two vacua on two sides are usually connected by a conjugate transformation of the discrete symmetry~\cite{Gelmini:2020bqg, Fu:2024jhu}. But for modular DWs, the two vacua at $ \tau  = {\rm i}$ and $\omega$, are not related by a modular transformation. The profile of energy density in the wall also turns out to be asymmetric along the coordinate perpendicular to the surface of the wall. Taking supergravity effects into account, in the background of a fixed dilaton field $S$, we show that the degeneracy may be lifted, leading to a bias term in the potential allowing DWs to collapse. We study the resulting gravitational wave (GW) spectra arising from the dynamics of such modular DWs, and assess their observability by current and future experiments, as a window into modulus stabilisation and modular flavour symmetry. 

The layout of the remainder of the paper is as follows. In Sec.~\ref{sec:modDW} we compute the modular DW solution and discuss its properties in the global supersymmetry limit. GWs from modular DWs are investigated in Sec.~\ref{sec:gw}. We summarise our main conclusions in Sec.~\ref{sec:sum}. In appendix~\ref{sec:appA}, we present the expressions for modular forms and modular-invariant functions used in this paper. The complete potentials and field equations in rigorous ${\cal N} = 1$ supergravity are shown in appendices~\ref{sec:appB} and \ref{sec:appC}, respectively.

\section{Modular symmetry and modular domain walls} \label{sec:modDW}
\begin{figure}[t!]
	\centering
    \includegraphics[width=0.48\linewidth]{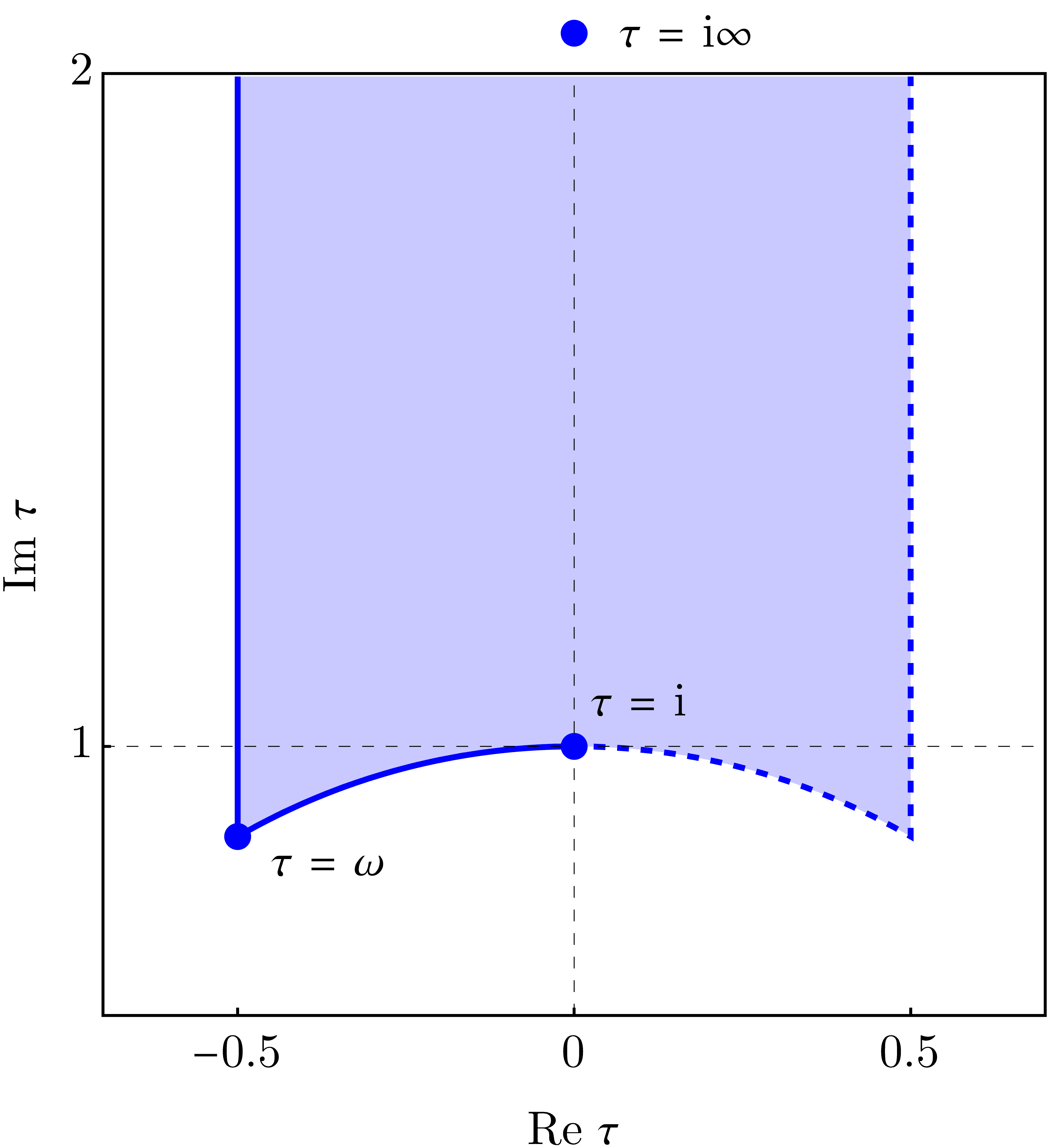}
	\caption{Fundamental domain of the modular group. We use blue dots to label the fixed points. }
	\label{fig:FDspec}
\end{figure}

String compactifications on specific manifolds are believed to exhibit modular symmetries, which leave the physical action invariant under modular transformations on moduli fields. The modulus field $\tau$ can be described by a complex parameter within the upper-half complex plane $\mathbb{C}_+$, transforming under an element $\gamma$ of the modular group $\overline{\Gamma} \cong {\rm PSL}(2,\mathbb{Z})$ as
\begin{eqnarray}
\gamma: \tau \rightarrow \dfrac{a \tau + b}{c \tau + d} \; ,
\label{eq:lintran}
\end{eqnarray}
with $a, b, c, d \in \mathbb{Z}$ and $a d - b c = 1$. $\overline{\Gamma}$ contains two generators, namely, the duality transformation $S$: $\tau \to -1/\tau$, and the shift transformation $T$: $\tau \to \tau + 1$. Acting all elements $\gamma$ on a given point in $\mathbb{C}_+$ generates an orbit of $\tau$. Then one can define the so-called fundamental domain, as shown in Fig.~\ref{fig:FDspec}, which intersects with each of these orbits at one and only one point. In the modular-invariant flavour models with a single modulus field, it is usually sufficient to scan $\tau$ within $\cal{G}$, since one can always send $\tau$ to $\cal{G}$ via a modular transformation, while keeping the physical observables unchanged~\cite{Novichkov:2018ovf}. Additionally, we have three fixed points within the fundamental domain, namely, $\I$, $\omega\equiv e^{2\pi\I/3}$, and $\I \infty$, at which residual symmetries $Z_2^S$, $Z_3^{ST}$ and $Z_N^T$ (with $N$ being the level of finite quotient groups) are respectively maintained after the spontaneous breakdown of modular symmetries. 

The VEV of $\tau$ can be determined by minimising the scalar potential constructed from supergravity theory. Here we consider heterotic orbifolds including a single K\"ahler modulus $\tau$ and a dilaton $S$, focusing primarily on the global supersymmetry limit of ${\cal N} = 1$ supergravity. The supergravity action is determined by a modular-invariant K\"ahler function $G(\varall)$ defined as~\cite{Cremmer:1982en}
\begin{equation}
    G(\varall) = \calK(\varall)/\plk^2 + \ln \left|\calW(\tau,S)/\plk^3 \right|^2 \; ,
    \label{eq:kahlerG}
\end{equation}
where $\calK(\varall)$ and $\calW(\tau,S)$ represent respectively the K\"ahler potential and superpotential, $\overline{\tau}$ and $\overline{S}$ denote the complex conjugates of $\tau$ and $S$, and $\plk = 2.43 \times 10^{18}~{\rm GeV}$ is the reduced Planck mass. 
The K\"ahler potential in our model contains two parts: one related to $S$ and the other to $\tau$. We assume that the K\"ahler potential for the K\"ahler modulus $\tau$ takes the minimal form, but we do not specify the exact form of the K\"ahler potential for $S$. As a result, the overall K\"ahler potential can be written as
\begin{equation}
   	{\cal K}(\varall) = - 3 \Lambda_K^2 \ln(2 \,{\rm Im}\,\tau) + \plk^2 K^{\rm dil}(S,\overline{S})   \; ,
    \label{eq:kahler-potential}
\end{equation}
where the factor 3 stems from three copies of two-tori in string compactifications~\cite{Cvetic:1991qm}, and $K^{\rm dil}(S,\overline{S})$ is a dimensionless real function of $S$ and $\overline{S}$. Note that $\tau$ and $S$ appearing hereafter are regarded as dimensionless. Their dimensions can be restored by including the scales of corresponding K\"ahler potentials. Specifically, the K\"ahler potential of $S$ is associated with $\plk$, whereas that of $\tau$ is associated with an energy scale $\Lambda_K$ which we assume to be much smaller than $\plk$. This ensures that we can take the global supersymmetry limit $\plk \to \infty$ for the K\"ahler modulus. Additionally, we assume that the dilaton is stabilised at a higher energy scale, independent of the K\"ahler modulus.\footnote{One example is the Kachru-Kallosh-Linde-Trivedi (KKLT) scenario~\cite{Kachru:2003aw}, where the dilaton superpotential ${\cal W}_{\rm KKLT} \propto \xi - B e^{-\zeta S}$ (with $\xi$, $\zeta$ and $B$ being constants). Such a superpotential, when combined with $K^{\rm dil} = -\ln(S + \overline{S})$, results in a scalar potential which has an minimum at $S \sim \ln(\xi/B\zeta)$.} Its VEV then leads to the 4d gauge coupling $g_4^2/2 = 1/\langle S + \overline{S} \rangle$. Nevertheless, the K\"ahler modulus $\tau$ can be stabilised via the gaugino condensation mechanism~\cite{Dine:1985rz, Nilles:1982ik, Ferrara:1982qs} at a relatively low scale, due to the following non-perturbative superpotential~\cite{Cvetic:1991qm}
\begin{equation}
    {\cal W}(\tau,S) = {\Lambda^3_W}H(\tau)\Omega(S) \; ,
    \label{eq:superp-para}
\end{equation}
where $\Lambda_W$ denotes the energy scale at which the gaugino condensation occurs, $\Omega(S)$ refers to a function of the dilaton $S$, and $H(\tau) = [j(\tau)-1728]^{m/2}j(\tau)^{n/3}$ with $j(\tau)$ being the modular-invariant Klein $j$-function and $m,n$ being non-negative integers.\footnote{We require that the function $H(\tau)$ do not introduce any singularity within $\cal{G}$. In the most general case, $H(\tau) \equiv [j(\tau)-1728]^{m/2}j(\tau)^{n/3} {\cal P}(j(\tau))$ with ${\cal P}(j(\tau))$ being a polynomial of $j(\tau)$. For simplicity, here we choose ${\cal P}(j(\tau)) = 1$.} We show the definition of $j(\tau)$ in appendix~\ref{sec:appA}. Eq.~(\ref{eq:superp-para}) implies that ${\cal W}(\tau,S)$ is modular invariant, as a consequence of the modular invariance of the K\"ahler function. 

The complete scalar potential in ${\cal N} = 1$ supergravity is given in Eq.~(\ref{eq:single-ponten-surga}) of appendix~\ref{sec:appB}. In the global supersymmetry limit, the scalar potential for the K\"ahler modulus reduces to
\begin{equation}
    V(\tau, \overline{\tau}) = \calK^{\tau\overline{\tau}}\partial_\tau \calW \partial_{\overline{\tau}}\overline{\calW} = \frac{\Lambda^4_V(2\,{\rm Im}\,\tau)^2_{}}{3}\left| H^\prime(\tau) \right|^2 \; .
    \label{eq:global-ponten}
\end{equation}
where $\partial_{\tau} \equiv \partial/\partial_\tau$, $\calK^{\tau\overline{\tau}} $ is the inverse of K\"ahler metric $\calK_{\tau\overline{\tau}} \equiv \partial_{\tau}\partial_{\overline{\tau}}\calK = 3\Lambda^2_K/(2\,{\rm Im}\,\tau)^2$, and the overall scale $\Lambda^4_V \equiv \Lambda_W^6 |\Omega|^2 e^{K^{\rm dil}}/\Lambda^2_{K}$ has been defined.\footnote{The factor $e^{K^{\rm dil}}$ comes from $e^{\calK/\plk^2}$ in Eq.~(\ref{eq:scalar-potential}). Substituting the K\"ahler potential $\plk^2 K^{\rm dil}(S,\overline{S})$ of the dilaton into $e^{K^{\rm dil}}$, we are left with this factor.} Since we assume $S$ is already fixed at its VEV, we can absorb $|\Omega|^2$ into $\Lambda_W$ and $e^{K^{\rm dil}}$ into $\Lambda_K$ without loss of generality. 

\begin{figure}[t!]
    \centering		
    \includegraphics[width=\textwidth]{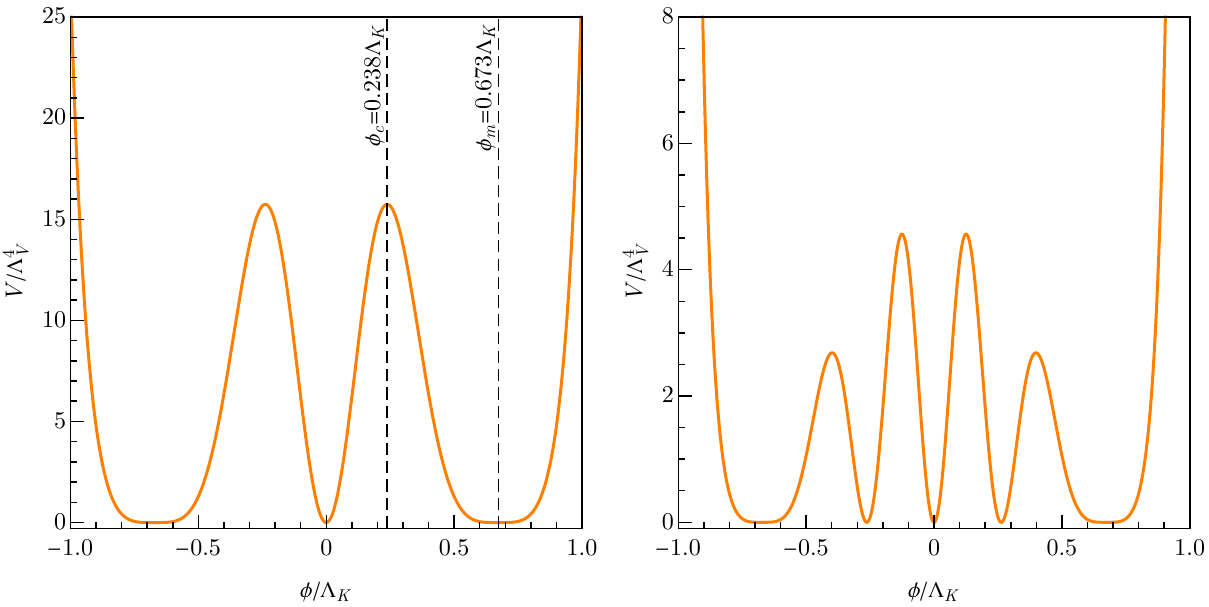} 	
    \caption{Projection of the scalar potential along the lower boundary of the fundamental domain (the unit arc). The field $\phi$ is defined as $\phi \equiv \Lambda_K \ln [\tan(\theta/2)]/\sqrt{2}$ with $\theta$ being the complex phase of $\tau = e^{\I \theta}$. In the {\it left} panel, we choose $H(\tau) = j(\tau)/1728$. The local maximum $\phi_c = 0.238\Lambda_K$ and the minimum $\phi_m = 0.673\Lambda_K$ which corresponds to $\tau = \omega$ are labelled by dashed lines. In the {\it right} panel, we choose $H(\tau) = j(\tau)[j(\tau)-1728]/1728^2$.}
    \label{fig:poten} 
\end{figure}

The stabilisation of $\tau$ by minimising the scalar potential obtained from gaugino condensation has been broadly explored in previous literature, see, e.g., Refs.~\cite{Ferrara:1989qb, Font:1990nt, Gonzalo:2018guu,Novichkov:2022wvg,Leedom:2022zdm,King:2023snq}. For the scalar potential $V(\tau, \overline{\tau})$ given in Eq.~(\ref{eq:global-ponten}), depending on the values of $m$ and $n$, the vacua of $V(\tau,\overline{\tau})$ can be classified into the following three cases:
\begin{enumerate}[label=(\alph*)]
    \item $\bm{m = n = 0}$: $V(\tau,\overline{\tau})$ is vanishing. The modulus $\tau$ is not fixed.
    \item $\bm{m = 0, \; n \neq 0}$ {\bf or} $\bm{m \neq 0, \; n = 0}$:  $V(\tau,\overline{\tau})$ has two Minkowski vacua at the fixed points $\tau = \I$ and $\tau = \omega$. 
    The scalar potential along the lower boundary of fundamental domain, i.e., the unit arc, is shown in the left panel of Fig.~\ref{fig:poten}.
    \item $\bm{m \neq 0, \; n \neq 0}$: Apart from $\tau = \I$ and $\omega$, there is an additional Minkowski vacuum on the lower boundary of ${\cal G}$. The scalar potential along the unit arc is shown in the right panel of Fig.~\ref{fig:poten}. 
\end{enumerate}

DW solutions that connect different vacua of $\tau$ may be obtained by solving the field equations of $\tau$. We focus on the static planar DW solution $\tau = \tau(z)$ with $z$ being the spatial coordinate transverse to the wall.  The complete set of first-order field equations within the framework of ${\cal N} = 1$ supergravity can be found in Refs.~\cite{Cvetic:1992bf,Cvetic:1996vr}, and also in appendix~\ref{sec:appC} of this paper. Generally speaking, if the gravitational effects are significant, the spacetime metric will undergo drastic variation along the $z$-direction, especially near the wall. However,  in the global supersymmetry limit under consideration, the variation of the spacetime metric is suppressed by $\plk$ (See appendix~\ref{sec:appC} for more details). Consequently,  the field equations can be written as~\cite{Cvetic:1992bf,Cvetic:1996vr} 
\begin{align}
    \frac{\partial \tau}{\partial z} & = e^{\I \Theta}\calK^{\tau \overline{\tau}}\frac{\partial \overline{\calW}}{\partial \overline{\tau}} \; , \label{eq:firsteof-tau} \\
    {\rm Im}\,\left(\frac{\partial_z \calW} {\calW} \right) &  =  \frac{\partial \Theta}{\partial z}  = 0 \; ,
    \label{eq:firsteof-theta}
\end{align}
with $\Theta$ being the complex phase of $\calW$, \footnote{More precisely, $\Theta$ determines the direction in field space along which the DW configuration minimises the energy. It can be seen from Eq.~\eqref{eq:firsteof-tau} that $\partial_z \calW = e^{\I \Theta}\calK^{\tau \overline{\tau}}|\partial_\tau \calW|^2 \propto e^{{\rm i} \Theta}$. Therefore, in the global supersymmetry limit $\calW(z)$ traces a straight line in the complex $\calW$-plane,  indicating that $\arg \calW(z) = \Theta$ remains constant. Nevertheless, $\Theta$ can vary with $z$ and is not always the phase of $\calW$ in supergravity.}
and $\overline{\calW}$ represents the complex conjugate of $\calW$. Eq.~(\ref{eq:firsteof-theta}) indicates that the phase $\Theta$ keeps invariant along the entire trajectory. Since $\calW(\tau)$ is a modular-invariant function, it is straightforward to see that the value of $\calW(\tau)$ is real at any point on the unit arc $|\tau | = 1$, given that the relations $\calW(\tau) = \calW(-1/\tau)$ and $\calW(\tau) = \overline{\calW}(-\overline{\tau})$ should be satisfied. Therefore, the unit arc precisely represents the DW trajectory which separates the two vacua, $\tau = \I$ and $\omega$, in the field space.

Along the unit arc, the modulus $\tau$ can be parametrised as $\tau = e^{\I \theta}$ where $\theta$ is the complex phase. As the kinetic Lagrangian ${\cal L}_{\rm kin} = 3\Lambda_K^2 \partial_\mu \theta \partial^\mu \theta /(4\sin^2\theta)$ in terms of $\theta$ is not canonical, it is more convenient to convert $\theta$ into a normalised field by defining $\phi \equiv \sqrt{3/2}\Lambda_K\ln[\tan(\theta/2)]$.  Since $\calW$ should be real along the $\phi$-direction, we can thereby rewrite Eq.~(\ref{eq:firsteof-tau}) as
\begin{equation}
    \frac{\partial \phi}{\partial z} = 2\frac{\partial \calW }{\partial \phi} \; .
    \label{eq:eof-phi-o1}
\end{equation}
Squaring both sides of the above equation and differentiating with respect to $\phi$, we come into the second-order field equation
\begin{equation}
    \frac{\partial^2 \phi}{\partial z^2 } = \frac{\partial V}{\partial \phi} \; ,
    \label{eq:eof-phi-o2}
\end{equation}
where $V = \calK^{\tau\overline{\tau}}\partial_\tau \calW \partial_{\overline{\tau}}\overline{\calW} = 2(\partial \calW / \partial \phi)^2$ has been adopted.

\begin{figure}[t!]
    \centering		
    \includegraphics[width=1\textwidth]{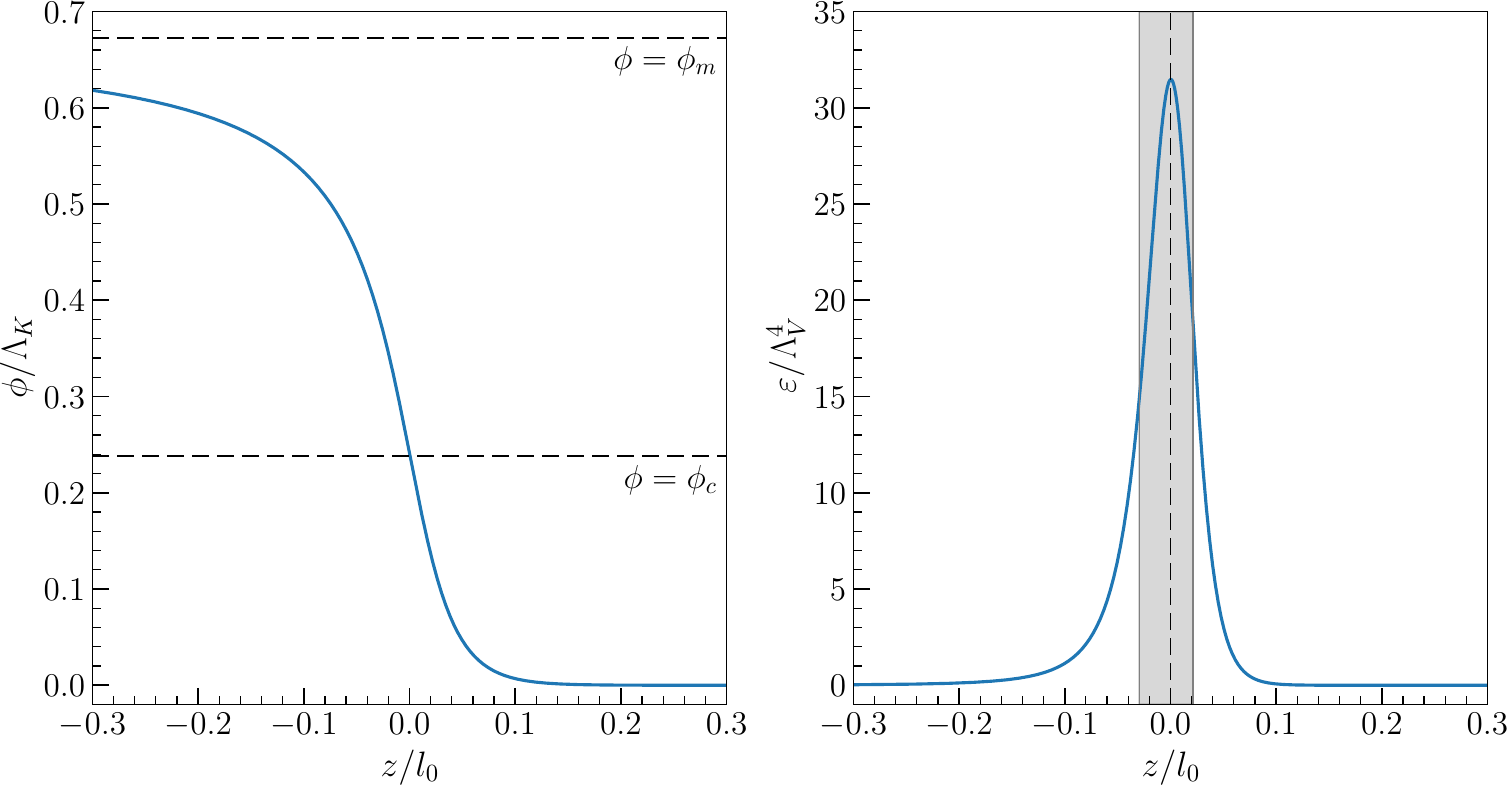} 	
    \caption{{\it Left}: The planar DW solution in {\it Case}~(b) where we choose $H(\tau) = j(\tau)/1728$. The convenient unit $l_0 \equiv \Lambda_K^2/\Lambda_W^3$ has been defined; {\it Right}: The distribution of energy density $\varepsilon$ along $z$-direction. The grey-shaded region corresponds to the thickness of DW.}
    \label{fig:dw_sol} 
\end{figure}

 As an illustrative example, we consider {\it Case} (b) with $m = 0$ and $n \neq 0$,\footnote{One can also choose $m \neq 0$ and $n = 0$. For example, one can choose $m = 2$ such that $H(\tau) = [j(\tau) - 1728]/1728$, which differs from the case where $H(\tau) = j(\tau)/1728$ by a constant. The scalar potential $V(\tau, \overline{\tau})$ remains unchanged as it only depends on $H^\prime(\tau)$.} while leaving {\it Case} (c) for future study.  More specifically, we choose $H(\tau) = j(\tau)/1728$ without loss of generality.\footnote{We choose $m=0$ and $n = 3$, and rescale $H(\tau)$ by a factor $j(\I) = 1728$.} Then we solve Eq.~(\ref{eq:eof-phi-o1}) numerically by choosing the initial condition $\phi(0) = \phi_c$ with $\phi_c = 0.238\,\Lambda_K$ corresponding to the barrier peak of $V(\tau, \overline{\tau})$. The result is shown in the left panel of \figref{fig:dw_sol}. One can observe that as $z \to \pm \infty$, $\phi$ tends to $0$ ($\tau = \I$) and $\phi_m = 0.673\,\Lambda_K$ ($\tau = \omega$), respectively. Compared to the limit at $\phi = \phi_m$, the approach to $\phi = 0$ is faster. This can be understood from the steeper increase in the potential from $\phi = 0$ to the top of the barrier, as shown in Fig.~\ref{fig:poten}. 
 
In the right panel of Fig.~\ref{fig:dw_sol}, we present the variation 
 of the energy density $\varepsilon = (\partial_z\phi)^2/2 +  2(\partial_\phi \calW )^2$ along the $z$-axis. We can continue to calculate the surface energy tension $\sigma$ of the DW, by integrating $\varepsilon$ in the $z$-direction, namely,
\begin{equation}
\begin{split}
    \sigma & = \int^\infty_{-\infty} {\rm d}z \left[ \frac{1}{2}\left(\partial_z\phi\right)^2 +  2\left(\partial_\phi \calW \right)^2\right] \\
    & = \int^\infty_{-\infty} {\rm d}z \, \frac{1}{2} \left( \partial_z \phi - 2  \partial_{\phi}\calW\right)^2 + 2| \Delta \calW | \; ,
    \end{split}
    \label{eq:tension}
\end{equation}
where $\Delta \calW \equiv \calW(\phi(z\to \infty)) - \calW(\phi(z \to -\infty))$.\footnote{Although Eq.~(\ref{eq:tension}) is derived from a specific model, $\sigma = 2|\Delta \calW|$ holds generally for supersymmetric DWs~\cite{Cvetic:1992bf}.} As $\partial_z \phi - 2\partial_{\phi}\calW = 0$ should be satisfied by Eq.~(\ref{eq:eof-phi-o1}), we can directly gain the surface tension as
\begin{equation}
    \sigma = 2|\Delta \calW| = 2\Lambda^3_W \; ,
    \label{eq:tension-result}
\end{equation}
where  $H(\I)=1$ and $H(\omega) = 0$ have been used. It is the gaugino condensation scale $\Lambda_W$ that solely determines the tension energy of modular DWs, which coincides with the ``no-scale'' behaviour of the K\"ahler potential given in Eq.~\eqref{eq:kahler-potential}.

Since the DW is asymmetric in our case, it is more convenient to define the partial thicknesses $\delta_\pm$ of DW separately on both sides of the peak of the energy barrier, which we have selected to be at $z = 0$, and the total thickness becomes $\delta = \delta_+ + \delta_-$. More concretely, the partial thicknesses $\delta_\pm$ can be determined by
\begin{equation}
    \begin{split}
         \int^0_{-\delta_-} {\rm d}z \, \varepsilon(z) = \, &  64.38\% \int^0_{-\infty} {\rm d}z \, \varepsilon(z) \; , \\
         \int^{\delta_+}_{0} {\rm d}z \, \varepsilon(z) = \, & 64.38\% \int^{\infty}_{0} {\rm d}z \, \varepsilon(z) \; ,
    \end{split}
\end{equation}
where we have drawn on the factor 64.38\% from the typical $Z_2$-symmetric DW~\cite{Fu:2024jhu}. Numerical results show that $\delta_- = 0.0294\,l_0$ and $\delta_+ = 0.0212\,l_0$ with $l_0 \equiv \Lambda_K^2/\Lambda_W^3$. In fact, we can also make an order-of-magnitude estimate of the DW thickness. Since the DW contains most of the energy within its thickness $\delta$, the surface tension can be estimated as $\sigma \sim (\phi^2_c/\delta^2 + V_{\rm max})\delta$ with $V_{\rm max}$ being the height of the potential barrier. As a result, we roughly have $\delta \sim \Lambda_W^3/V_{\rm max} \sim {\cal O}(10^{-2})\,l_0$, coinciding with the numerical result.

At the end of this section, let us emphasise the difference between the modular DWs considered here and the DWs arising from the spontaneous breaking of discrete symmetries in field theory. As we have seen, the presence of degenerate vacua originates from the properties of modular forms, rather than as a consequence of an explicit discrete symmetry in the scalar potential. For DWs in field theory, the two vacua on two sides are usually connected by a conjugate transformation of a certain discrete symmetry. But for modular DWs, we cannot relate the two vacua, $\tau = \I$ and $\tau = \omega$, through a modular transformation. This explains why the DW solution obtained in our model is asymmetric. It is interesting to point out that modular transformations can indeed relate $\tau$ in different domains. For instance, if $\tau = \I$ corresponds to  a vacuum, we will obtain an infinite number of degenerate vacua at $\tau = \I + l$ (with $l$ being an integer). However, all these vacua give rise to the same value of $\calW$. According to Eq.~(\ref{eq:tension-result}), the surface tension energy $\sigma$ of the modular DW depends only on the difference in the values of $\calW$ at different vacua, hence no DW solution with non-zero $\sigma$ can be obtained between different domains of the modular group~\cite{Cvetic:1992bf}.

\section{Gravitational waves from modular domain walls} \label{sec:gw}

As discussed in the previous section, the scalar potential given in Eq.~\eqref{eq:global-ponten}  leads to two degenerate vacua at $\tau = \I$ and $\tau = \omega$, enabling the formation of stable DWs. The energy density of DWs quickly comes to dominate our Universe, conflicting with current cosmic microwave background (CMB) observations. In order to prevent longevity of DWs, we allow them to annihilate by introducing a bias term in the scalar potential that breaks the vacua degeneracy. Interestingly, if we slightly relax the global supersymmetry limit $\plk \to \infty$, the dilaton term in the rigorous  supergravity potential given in Eq.~\eqref{eq:single-ponten} can serve as a small bias term suppressed by $\plk$. To be more specific, with this term added back, the scalar potential can be rewritten as
\begin{equation}
    V(\tau, \overline{\tau}) = \frac{\Lambda^6_W}{\Lambda^2_K}\frac{(2\,{\rm Im}\,\tau)^2_{}}{3}\left| H^\prime(\tau) \right|^2  + \frac{\Lambda^6_W}{\plk^2}\calA|H(\tau)|^2\; ,
    \label{eq:ponten-bias}
\end{equation}
where the factor $\calA$  comes essentially from the $F$-term of $S$, and can be expressed as
\begin{equation}
    \mathcal{A} = \frac{|\Omega^{}_S+ K^{\rm dil}_S\Omega|^2_{}}{ K^{\rm dil}_{S\overline{S}}|\Omega|^2_{}} \; ,
    \label{eq:AS}
\end{equation}
with $\Omega^{}_S \equiv \partial \Omega / \partial S$, $K_S^{\rm dil} \equiv \partial K^{\rm dil}_{}/\partial S$ and $K_{S\overline{S}}^{\rm dil} \equiv \partial^2_{} K^{\rm dil}_{}/(\partial S\partial \overline{S})$ being defined. $H(\tau)$ takes different values at $\tau = \I$ and $\omega$, namely, $H(\I)=1$ and $H(\omega)=0$. Therefore, the potential energy difference between two vacua provides a bias term
\begin{eqnarray}
    V_{\rm bias} = V|_{\tau = \I} - V|_{\tau = \omega} = \frac{\Lambda^6_W}{\plk^2}\calA \; .
    \label{eq:Vbias}
\end{eqnarray}
The presence of $V_{\rm bias}$ imposes an additional pressure force on the wall, causing the false vacuum to shrink. As a consequence, DWs start to collapse, generating a characteristic stochastic GW spectrum~\cite{Vilenkin:1981zs, Gelmini:1988sf, Larsson:1996sp, Hiramatsu:2013qaa, Hiramatsu:2012sc, Saikawa:2017hiv,Roshan:2024qnv}. We consider the scenario where the annihilation occurs in the radiation-dominated era, indicating that the annihilation temperature should be smaller than the reheating  temperature. The annihilation temperature is given by~\cite{Saikawa:2017hiv}
\begin{equation}
    T^{}_{\rm ann} = 9.46 \times 10^{17}_{}~{\rm GeV}~C^{-1/2}_{\rm ann}\mathcal{S}^{-1/2}_{}\left( \frac{g_*(T_{\rm ann})}{100}\right)^{-1/4} \left(\frac{\sigma}{\plk^3}\right)^{-1/2}_{} \left(\frac{V_{\rm bias}}{\plk^4}\right)^{1/2}  \; ,
    \label{eq:anntemp}
\end{equation}
where $g_*(T_{\rm ann})$ is the relativistic degrees of freedom for the radiation energy density at the temperature $T_{\rm ann}$. In the majority of the parameter space in our model, $T_{\rm ann} \gtrsim 1~{\rm GeV}$, hence we approximately have $g_*(T) \simeq 100$.  Additionally, we set $\mathcal{S}=0.8$ for the area parameter and $C_{\rm ann}=2$ for the dimensionless annihilation constant, as an analogy to the $Z_2$-symmetric DW case~\cite{Kawasaki:2014sqa}. The presence of the asymmetry in Fig.~\ref{fig:dw_sol} may result in a non-uniform annihilation time of the DW network. However, the values of $\delta_-$ and $\delta_+$ indicate that the level of asymmetry is of $\sim{\cal O}(30\%)$, which is not large enough to significantly
affect the phenomenology, given the present level of accuracy of the theoretical calculations and expected experimental uncertainty. However, in the future, it may become observable and provide a distinctive signature of this approach.

\begin{figure}[t!]
    \centering		
    \includegraphics[width=0.7\textwidth]{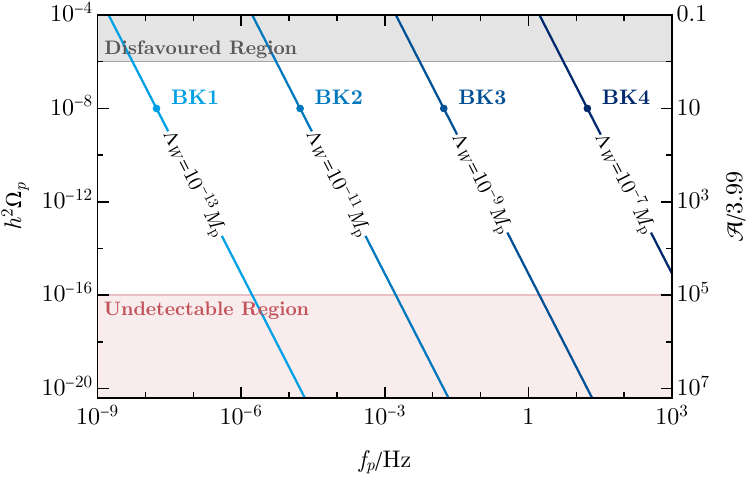}	
    \caption{Correlations between the peak frequency $f_p$ and the peak amplitude $\Omega_p h^2$ of GW spectra for $\Lambda_W = 10^{-13}\plk$, $10^{-11}\plk$, $10^{-9}\plk$, and $10^{-7}\plk$. The grey-shaded disfavoured region stems from the measurements on the number of extra neutrino species~\cite{Smith:2006nka, Sendra:2012wh, Pagano:2015hma, Gouttenoire:2019kij}. The red-shaded region is unlikely to be detectable by any GW observatories in the foreseeable future.}
    \label{fig:gw_det} 
\end{figure}

\begin{figure}[t!]
    \centering		
\includegraphics[width=0.7\textwidth]{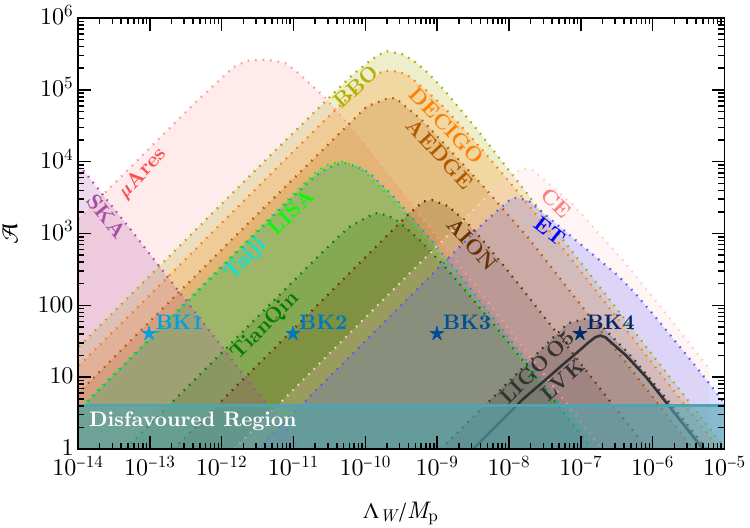}	
    \caption{The parameter space of $\{\Lambda_W, \calA \}$ corresponding to $\varrho \leq 10$ for various GW detectors. The region disfavoured by measurements of the number of extra neutrino species is shown in teal shading. We also mark four {\bf BK} points with pentagrams.}
    \label{fig:constraints} 
\end{figure}

The peak frequency $f_p$ and peak energy density amplitude $\Omega_p h^2$ of GWs from DWs can be estimated as~\cite{Saikawa:2017hiv}
\begin{equation}
\begin{split}
f_p^{} & \simeq 1.53 \times 10^{11}_{}~{\rm Hz} ~C^{-1/2}_{\rm ann} \mathcal{S}^{-1/2}_{} \left( \frac{g_*(T_{\rm ann})}{100}\right)^{-1/12} \left(\frac{\sigma}{\plk^3}\right)^{-1/2}_{} \left(\frac{V_{\rm bias}}{\plk^4}\right)^{1/2} \; , \\ 
\Omega^{}_{p} h^2_{} & \simeq 8.63 \times 10^{-7}_{} \epsilon \mathcal{S}^4_{} C^2_{\rm ann} \left( \frac{g_*(T_{\rm ann})}{100}\right)^{-1/3} \left(\frac{\sigma}{\plk^3}\right)^{4}_{} \left(\frac{V_{\rm bias}}{\plk^4}\right)^{-2}  \; ,
\end{split}
\label{eq:peak-a}
\end{equation}
where the effective relativistic degrees of freedom at the present time $g_{*0} = 3.36$ and $g_{*s0} = 3.91$ have been adopted, and the efficiency parameter is taken to be $\epsilon \simeq 0.7$~\cite{Hiramatsu:2012sc}. Substituting the expressions of $\sigma$ and $V_{\rm bias}$ into Eqs.~(\ref{eq:anntemp}) and (\ref{eq:peak-a}), we could roughly have the following relations
\begin{equation}
\begin{split}
    T_{\rm ann} &\simeq 5.29 \times 10^{17}~{\rm GeV}~\sqrt{\calA } \left(\Lambda_W / \plk\right)^{3/2} \; , \\
    f_p &\simeq 8.54 \times 10^{10}~{\rm Hz}~\sqrt{\calA } \left(\Lambda_W / \plk\right)^{3/2} \; , \\
    \Omega_{p} h^2 &\simeq 1.59 \times 10^{-5}\calA^{-2} \; ,
\end{split}
\label{eq:peak-range}
\end{equation}
where one could see that $\Omega_{p} h^2$ depends only on $\calA$, whereas $T_{\rm ann}$ and $f_p$ depend on both $\calA$ and $\Lambda_W$. Apparently, we need $\Lambda_W/M_p \ll 1$ to avoid extremely large values of $T_{\rm ann}$ and $f_p$. Moreover, once the value of $\Lambda_W$ is fixed, $\Omega_{p} h^2$ and $f_p$ satisfy a power-law relation $\Omega^{}_{p} h^2_{} = 8.47\times 10^{38}_{}(\Lambda_W / \plk)^6 (f_p^{}/{\rm Hz})^{-4}$. We explicitly exhibit the correlations between $f_p$ and $\Omega^{}_{p}$ in Fig.~\ref{fig:gw_det} by choosing $\Lambda_W = 10^{-13}\plk$, $10^{-11}\plk$, $10^{-9}\plk$, and $10^{-7}\plk$, respectively. The energy density of a GW background in the early Universe decays as relativistic degrees of freedom, which should be constrained by the measurements on the number of extra neutrino species $\Delta N_{\rm eff}$~\cite{Caprini:2018mtu}. Combined analyses using Planck and other data indicate that the energy density of GW backgrounds produced before Big-Bang Nucleosynthesis (BBN) satisfies $h^2\Omega_{\rm GW} \lesssim 10^{-6}$~\cite{Smith:2006nka, Sendra:2012wh, Pagano:2015hma,Gouttenoire:2019kij}. We denote the region where $h^2\Omega_{\rm GW} \geq 10^{-6}$ as the ``Disfavoured Region'' in Fig.~\ref{fig:gw_det}. On the other hand, GWs with $h^2\Omega_{\rm GW}  \lesssim 10^{-16}$ are unlikely to be detectable by any current or planned experiments in the foreseeable future. We label this region as the ``Undetectable Region''. In order for the GWs from DWs to be testable, $\calA$ should be approximately within the range $4 \lesssim \calA \lesssim 4 \times 10^5$. As for the gaugino condensation scale, $\Lambda_W \lesssim 10^{-7}\plk$ corresponds to the peak frequencies $f_p \lesssim 10^3~{\rm Hz}$. The typical value of $\Lambda_W$ motivated by string theory can be as large as $\Lambda_W \sim 10^{13}~{\rm GeV} \sim 10^{-5}\plk$, which can keep the gravitino mass $m_{3/2}$ in the TeV range~\cite{Nilles:1990zd}. As we will see in Fig.~\ref{fig:constraints},  such a value of $\Lambda_W$ may just barely reach the edge of the detectable region. However, $\Lambda_W$ could also be smaller, depending on the dynamics of the hidden sector gauge group~\cite{Binetruy:1996uv,Binetruy:1995ta,Saririan:1996mr}. In this paper, we treat $\Lambda_W$ as a free parameter that can vary over a wide range.

\begin{figure}[t!]
    \centering		
\includegraphics[width=0.7\textwidth]{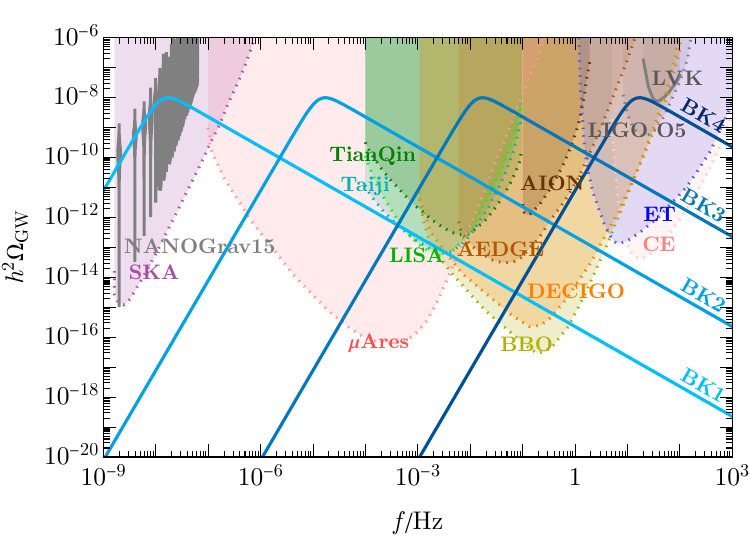}	
    \caption{Four benchmark GW spectra corresponding to {\bf BK1}---{\bf BK4} in Fig.~\ref{fig:gw_det},  together with the power-law-integrated sensitivity (PLIS) curves~\cite{Thrane:2013oya,Schmitz:2020syl, Breitbach:2018ddu} from SKA~\cite{Janssen:2014dka}, $\mu$Ares~\cite{Sesana:2019vho}, LISA~\cite{LISA:2017pwj}, Taiji~\cite{Ruan:2018tsw}, TianQin~\cite{TianQin:2015yph}, DECIGO~\cite{Kawamura:2020pcg}, BBO~\cite{Corbin:2005ny}, AEDGE~\cite{AEDGE:2019nxb}, AION km~\cite{Badurina:2019hst}, ET~\cite{Punturo:2010zz}, CE~\cite{Reitze:2019iox}, and aLIGO~\cite{LIGOScientific:2014pky}. In addition, we show the recent NANOGrav 15-year results~\cite{NANOGrav:2023gor} and the excluded region by LIGO-Virgo-Kagra (LVK) Run O3~\cite{KAGRA:2021kbb}.}
    \label{fig:gw_sensty} 
\end{figure}

We select four benchmark ({\bf BK}) points with $\calA = 40$ and $\Lambda_W/\plk = \{10^{-13}$, $10^{-11}$, $10^{-9},10^{-7} \}$, and estimate the corresponding GW spectra using the following broken power-law parametrisation~\cite{Caprini:2019egz, NANOGrav:2023hvm}
\begin{eqnarray}
h^2_{} \Omega^{}_{\rm GW} = h^2_{} \Omega^{}_p \frac{(r+s)^u}{\left(s x^{-r / u}+r x^{s / u}\right)^u} \ ,
\label{eq:spec-par}
\end{eqnarray}
where $x \equiv f/f_p$, and the parameters $r = 3$, $s = u =1$ are chosen~\cite{Hiramatsu:2013qaa}. Our results are shown in Fig.~\ref{fig:gw_sensty}. For comparison, we also present the power-law-integrated sensitivity (PLIS) curves~\cite{Thrane:2013oya,Schmitz:2020syl, Breitbach:2018ddu} from the {\it Square Kilometre Array} (SKA)~\cite{Janssen:2014dka}, $\mu$Ares~\cite{Sesana:2019vho}, the {\it Laser
Interferometer Space Antenna} (LISA)~\cite{LISA:2017pwj}, Taiji~\cite{Ruan:2018tsw}, TianQin~\cite{TianQin:2015yph}, the {\it Deci-hertz Interferometer Gravitational-wave Observatory} (DECIGO)~\cite{Kawamura:2020pcg}, the {\it Big-Bang Observer} (BBO)~\cite{Corbin:2005ny}, the {\it Atomic Experiment for Dark Matter and Gravity Exploration in Space} (AEDGE)~\cite{AEDGE:2019nxb}, the {\it Atom Interferometer Observatory and Network} (AION) km~\cite{Badurina:2019hst}, the {\it Einstein Telescope} (ET)~\cite{Punturo:2010zz}, the {\it Cosmic Explorer} (CE)~\cite{Reitze:2019iox}, the {\it Advanced Laser Interferometer Gravitational-Wave Observatory} (aLIGO)~\cite{LIGOScientific:2014pky}, the {\it Advanced Virgo} (aVirgo)~\cite{VIRGO:2014yos}, and the {\it Kamioka Gravitational Wave Detector} (KAGRA)~\cite{KAGRA:2018plz}. The results from the LIGO-Virgo-Kagra (LVK) third observing run (O3), combined with the earlier O1 and O2 runs, placed upper limits on the strength of the GW background within the frequency band $20~{\rm Hz} \leq f \leq 90.6~{\rm Hz}$ at the 95\% credible level~\cite{KAGRA:2021kbb}, which is shown by the solid grey curve in Fig.~\ref{fig:gw_sensty}. Recently, several pulsar timing array (PTA) projects, such as the {\it North American Nanohertz Observatory for Gravitational Waves} (NANOGrav)~\cite{NANOGrav:2023gor, NANOGrav:2023hvm}, the European PTA~\cite{EPTA:2023fyk,EPTA:2023xxk}, the Parkes PTA~\cite{Reardon:2023gzh} and the Chinese PTA~\cite{Xu:2023wog}, reported strong evidence of a stochastic GW background with the frequency around $10^{-8}~{\rm Hz}$. In Fig.~\ref{fig:gw_sensty}, we also exhibit the NANOGrav 15-year results using grey blobs. Explaining the NANOGrav 15-year data requires $\Lambda_W \lesssim 10^{-13}\plk$.

We can further estimate the capability of GW detectors to test the parameter space of our model in a more qualitative way
by calculating the signal-to-noise ratio (SNR)~\cite{Maggiore:1999vm,Allen:1997ad}
\begin{equation}
    \varrho=\left[n_{\mathrm{det}} t_{\mathrm{obs}} \int_{f_{\min }}^{f_{\max }} d f\left(\frac{\Omega_{\text {signal }}(f)}{\Omega_{\text {noise}}(f)}\right)^2\right]^{1 / 2} \; ,
    \label{eq:SNR}
\end{equation}
where $n_{\rm det} = 1$ for auto-correlated detectors and  $n_{\rm det} = 2$ for cross-correlated detectors, $t_{\rm obs}$ represents the observational time, $\Omega_{\text {signal}}$ represents the GW signal predicted in our model, following the form given in Eq.~(\ref{eq:spec-par}), and $\Omega_{\text {noise }}$ denotes the noise spectrum expressed in terms of the GW energy density spectrum which we directly take from the experiments mentioned above. 
Integrating $(\Omega_{\rm signal}/\Omega_{\rm noise})^2$ over the sensitive frequency ranges of individual GW detectors, we obtain the SNRs for given values of $\Lambda_W$ and $\calA$. In Fig.~\ref{fig:constraints}, We use different colour-shaded regions to indicate the parameter space of $\{\Lambda_W, \calA \}$ corresponding to $\varrho \leq 10$ for various GW detectors, which delineate the parameter ranges detectable by these detectors. We find that the gaugino condensation scale may be probed by multiple GW observatories across a broad frequency range.  More concretely, the ground-based laser interferometers like LIGO can touch the parameter range with $\Lambda_W \lesssim 10^{-5}\plk$. Results from the LVK O3 run have already excluded the region $10^{-8}\plk \lesssim \Lambda_W \lesssim 10^{-6}\plk$ and $\calA \lesssim {\cal O}(10)$. The space-based detectors, such as LISA, Taiji and Tianqin, are sensitive to $\Lambda_W \sim 10^{-10} M_{\rm p}$. PTA experiments, which are capable of detecting nano-Hz GWs, hold the potential to probe the parameter region where $\Lambda_W \lesssim 10^{-11}\plk$.

Before closing this section, let us make two additional comments on the values of $\calA$ and $\Lambda_K$. First, one may wonder whether large values of $\calA$ like $\calA \sim 10^{5}$ are theoretically reasonable. Here, instead of constructing a specific model for the dilaton stabilisation, we briefly demonstrate that large values of $\calA$ are feasible based on some empirical  hypotheses. Assuming the tree-level K\"ahler potential  $K^{\rm dil} = -\ln(S + \overline{S})$ and an exponential function $\Omega(S) = B e^{-\zeta^\prime S}$ (with $\zeta^\prime$ being an constant), we could identify from Eq.~(\ref{eq:AS}) that $\calA \simeq \zeta^{\prime2}\langle S + \overline{S} \rangle^2$ in the large $S$ limit. In this regard, a sizeable $\calA$ can be achieved by choosing a sufficiently large $S$. Keeping in mind that the 4d gauge coupling $g_4^2/2 = 1/\langle S + \overline{S} \rangle$, a large $\calA$ would correspond to a small gauge coupling constant. Second, $\Lambda_K$ is a completely unconstrained scale in our previous discussions, as we have already seen that neither $\sigma$ nor $V_{\rm bias}$ depends on $\Lambda_K$. However, the scalar potential $V(\tau,\overline{\tau})$ indeed depends on $\Lambda_K$. According to the percolation theory, in order for the DWs to be formed, the height of the potential barrier $V_{\rm max}$ and the bias term $V_{\rm bias}$ should satisfy~\cite{Saikawa:2017hiv}
\begin{equation}
    \frac{V_{\rm bias}}{V_{\rm max}} < \ln\left(\frac{1-p_c}{p_c}\right) = 0.795 \; ,
    \label{eq:percolation}
\end{equation}
where $p_c = 0.311$ is the critical probability above which an infinite cluster of the false vacuum appears in the space, supposing the system is treated as a three dimensional cubic lattice~\cite{Stauffer:1978kr}. Eq.~(\ref{eq:percolation}) results in a constraint on $\Lambda_K$ depending on $\calA$ as $\Lambda_K < 3.54\calA^{-1/2}\plk$.

\section{Summary} \label{sec:sum}
Modulus stabilisation is not only a key topic in string theory research but also plays a crucial role in low-energy modular flavour symmetries. It provides a dynamical mechanism for determining the modulus parameter, which dictates the flavour mixing structure of quarks and leptons. However, modulus stabilisation generally takes place at a very high energy scale, posing significant challenges for developing effective approaches to probe it. In this paper, we have investigated modular DWs and their testability in GW observatories in a class of modular-invariant supersymmetric models. In such models, a single K\"ahler modulus field $\tau$ can be stabilised at or near certain fixed points, $\tau = \I$ and $\tau = \omega$, due to the potential induced by gaugino condensation. In the global supersymmetry limit $\plk \to \infty$, such fixed points may be degenerate, leading to the formation of modular DWs in the early Universe. 

We have numerically computed the DW solution in our model and discussed in detail the properties of modular DWs. Their tension, i.e., mass per square unit, is of order $\Lambda_W^3$, where $\Lambda_W$ is the superpotential scale resulting from gaugino condensation. The modular DWs are very different from those arising from ordinary discrete symmetry breaking in field theory. Specifically, the two vacua at $ \tau  = {\rm i}$ and $\omega$ cannot be related by a modular transformation. As a result, the profile of energy density in the wall is asymmetric along the coordinate perpendicular to the surface of the wall. Although modular DWs are very distinct from ordinary DWs, we have found a consistent way to define their thickness. The total thickness of the walls is of order $10^{-2} \Lambda_K^2 / \Lambda_W^3$ where $\Lambda_K$ is the scale appearing in the K\"ahler potential of $\tau$. 

We also have shown that, taking supergravity effects into account, in the background of a fixed dilaton field $S$, the degeneracy between two vacua at $\tau = \I$ and $\omega$ may be lifted, leading to a bias term in the potential allowing the DWs to collapse. We have studied the resulting GW spectra arising from the dynamics of such modular DWs, parametrised by the gaugino condensation scale $\Lambda_W$ and a dimensionless dilaton term $\calA$, and assessed its observability by current and future experiments. In order for the GWs from DWs to be testable, $\calA$ should be approximately within the range $4 \lesssim \calA \lesssim 4 \times 10^5$. Values of $\Lambda_W$ within the range $10^{-14}\plk \lesssim \Lambda_W \lesssim 10^{-5}\plk$ are promising for examination by several GW detectors across a wide frequency range. We have identified a benchmark GW spectrum with $\Lambda_W = 10^{-13}\plk$ and $\calA = 40$ that fits the recent NANOGrav 15-year results very well.

It is worth mentioning that, although we have considered only a simplified model in this paper, modular DWs and their collapse may be a common feature in scalar potentials with modular symmetry. Extending our work to other string theory models or within the rigorous framework of supergravity would be interesting. It is also worthwhile to investigate DWs separating CP-violating vacua that are not precisely located on the boundary of the fundamental domain.

\acknowledgments

We would like to thank Nicole Righi and Masahito Yamazaki for useful discussions. XW would also like to thank João Penedo for helpful email exchanges. SFK acknowledges the STFC Consolidated Grant
ST/X000583/1 and the European Union's Horizon 2020 Research and Innovation programme under Marie Sklodowska-Curie grant agreement HIDDeN European ITN project (H2020-MSCA-ITN-2019//860881-HIDDeN).  XW acknowledges the Royal Society as the funding source of the Newton International Fellowship.
YLZ is partially supported by the National Natural Science Foundation of China (NSFC) under Grant Nos. 12205064, 12347103, and Zhejiang Provincial Natural Science Foundation of China under Grant No. LDQ24A050002.

\appendix
\section{Some useful modular forms and modular-invariant functions}\label{sec:appA}
The modular form $f(\tau)$ is a holomorphic function in terms of $\tau$ transforming under the modular group as $f\left(\gamma\tau\right)=(c \tau+d)^{k} f(\tau)$, with $k$ being its weight. Modular forms can be generated by the Dedekind $\eta$ function, a modular form with a weight of $1/2$, which is defined as
\begin{equation}
 \eta(\tau)\equiv q^{1 / 24} \prod_{n=1}^{\infty}\left(1-q^n\right) \; ,
\label{eq:dedeta}
\end{equation}
where $q = e^{2\pi {\rm i}\tau}_{}$. One can express $\eta(\tau)$ as the following $q$-expansions
\begin{equation}
\eta=q^{1 / 24}\left(1-q-q^2+q^5+q^7-q^{12}-q^{15}+\mathcal{O}\left(q^{22}\right)\right) \; .
\label{eq:etaexpan}
\end{equation}

The Eisenstein series $G^{}_{2k}(\tau)$ ($k>1$) are holomorphic modular forms with weights of $2k$, the definitions of which are
\begin{equation}
    G_{2 k}(\tau)=\sum_{\substack{n_1, n_2 \in \mathbb{Z} \\\left(n_1, n_2\right) \neq(0,0)}}\left(n_1+n_2 \tau\right)^{-2 k} \; .
    \label{eq:eisenstein}
\end{equation}
If $k=1$, we obtain the Eisenstein series $G^{}_2(\tau)$, which can be related to the Dedekind $\eta$-function via 
\begin{equation}
    \frac{\eta^\prime_{} (\tau)}{\eta(\tau)} = \frac{\rm i}{4\pi} G_2(\tau) \; .
\label{eq:hol-eisen}
\end{equation}
Notice that $G^{}_2(\tau)$ is a holomorphic function but not a modular form. Instead, one can define a non-holomorphic Eisenstein series of weight two as
\begin{equation}
    \widehat{G}_2(\tau) = G_2(\tau) - \frac{\pi}{{\rm Im}\,\tau} \; .
    \label{eq:nonholoG2}
\end{equation}

With the help of $\eta(\tau)$ and $G^{}_4(\tau)$, one can define a modular-invariant function which is called the Klein $j$-function as
\begin{eqnarray}
    j(\tau) = \frac{3^6 5^3}{\pi^{12}} \frac{G_4(\tau)^3}{\eta(\tau)^{24}} \; .
    \label{eq:Klein}
\end{eqnarray}
One can also express $j(\tau)$ in terms of the Dedekind $\eta$-function and its derivatives as
\begin{equation}
j = \left( \frac{72}{\pi^2}\frac{ \eta \eta'' - 3 \eta'^2}{\eta^{10}} \right)^3 = \left[ \frac{72}{\pi^2 \eta^6} \left( \frac{\eta'}{\eta^3} \right)' \right]^3 \; .
\end{equation}

\section{The scalar potential in ${\cal N } = 1$ supergravity} \label{sec:appB}
In this appendix we present the generic form of the moduli scalar potential in the rigorous ${\cal N} = 1$ supergravity. Again, we start from the K\"ahler function
\begin{equation}
    G(\varall) = \calK(\varall)/\plk^2 + \ln \left|\calW(\tau,S)/\plk^3 \right|^2 \; .
    \label{eq:kahlerG-app}
\end{equation}
The K\"ahler potential still takes the form given by Eq.~(\ref{eq:kahler-potential}), but we do not restrict ourselves to the global supersymmetry limit $\Lambda_K \ll \plk$. In general, one can define a ratio as $\fm \equiv \Lambda_K^2/\plk^2$. Under the modular transformation, we have ${\rm Im}\,\tau \to |c\tau + d|^{-2}_{} {\rm Im}\,\tau$. Therefore the modular invariance of the K\"ahler function indicates that the superpotential should possess a weight of $-3\fm$. Consequently, we have the following superpotential~\cite{Cvetic:1991qm}
\begin{equation}
    {\cal W}(\tau,S) = {\Lambda^3_W}\frac{H(\tau)\Omega(S)}{\eta^{6\fm}(\tau)} \; .
    \label{eq:superp-para-app}
\end{equation}
If $\Lambda_K$ and $\plk$ are precisely equal, $\fm = 1$, leading to the superpotential form given in Refs.~\cite{Leedom:2022zdm,King:2023snq}. Conversely, if 
$\Lambda_K \ll \plk$, $\fm = 0$, and Eq.~(\ref{eq:superp-para-app}) reduces to the  global supersymmetry limit of the superpotential, as shown in Eq.~(\ref{eq:superp-para}).

With the help of the K\"ahler function $G$, the scalar potential can be determined by~\cite{Cremmer:1982en} 
\begin{equation}
    V = \plk^4 e^{G}(\calK^{i\overline{j}}D_i G D_{\overline{j}} G-3) \; ,
    \label{eq:def-scalarpotential}
\end{equation}
where the covariant derivatives $D_i \equiv \partial^{}_i + (\partial_i {\cal K})/\plk^2$ are defined. Substituting the expression of $G$ into the above equation, we arrive at~\cite{Cremmer:1982en}
\begin{equation}
    V=e^{{\cal K}/\plk^2}\left({\cal K}^{i \bar{j}}_{} D^{}_i {\cal W} D^{}_{\bar{j}} {\cal \overline{W}}-3 |{\cal W}|^2_{}/\plk^2\right) \; ,
    \label{eq:scalar-potential}
\end{equation}
where $i,j = \tau, S$. Plugging Eqs.~(\ref{eq:kahler-potential}) and (\ref{eq:superp-para-app}) into Eq.~(\ref{eq:scalar-potential}), we can obtain the expression for the scalar potential as
\begin{equation}
    V(\tau, \overline{\tau}) = \frac{\Lambda^4_V }{(2\,{\rm Im}\,\tau)^{3\fm}_{}|\eta(\tau)|^{12\fm}_{}}\left[\frac{(2\,{\rm Im}\,\tau)^2_{}}{3}\left|\I H^\prime(\tau) + \frac{3\fm}{2\pi}\widehat{G}_2(\tau,\overline{\tau})H(\tau)\right|^2_{} + \fm\left(\calA -3 \right)|H(\tau)|^2\right] \; ,
    \label{eq:single-ponten}
\end{equation}
where the overall scale $\Lambda^4_V \equiv \Lambda_W^6 |\Omega|^2 e^{K^{\rm dil}}/\Lambda^2_{K}$, and $\calA$ depends on $S$ and $\overline{S}$ as
\begin{equation}
    \mathcal{A} = \frac{|\Omega^{}_S+ K^{\rm dil}_S\Omega|^2_{}}{ K^{\rm dil}_{S\overline{S}}|\Omega|^2_{}} \; ,
    \label{eq:AS-app}
\end{equation}
with $\Omega^{}_S \equiv \partial \Omega / \partial S$, $K_S^{\rm dil} \equiv \partial K^{\rm dil}_{}/\partial S$ and $K_{S\overline{S}}^{\rm dil} \equiv \partial^2_{} K^{\rm dil}_{}/(\partial S\partial \overline{S})$ being defined. $\calA$ essentially comes from the $F$-term of $S$ in the scalar potential, as we can see that $D^{}_S \calW = \Lambda^3_W H(\tau) (\Omega^{}_S + K^{\rm dil}_{S}\Omega) / \eta^{6\fm}_{}(\tau)$. 

With the help of Eq.~(\ref{eq:single-ponten}), we can compare the scalar potential in the global supersymmetry limit with that in supergravity. If $\fm = 0$, one can get
\begin{equation}
    V(\tau, \overline{\tau}) = \frac{\Lambda^4_V (2\,{\rm Im}\,\tau)^2_{}}{3}\left| H^\prime(\tau) \right|^2 \; ,
    \label{eq:global-ponten-add}
\end{equation}
which is precisely the scalar potential in the global supersymmetry limit shown in Eq.~(\ref{eq:global-ponten}). On the contrary, if we set $\fm = 1$, the scalar potential turns out to be
\begin{equation}
    V(\tau, \overline{\tau}) = \frac{\Lambda^4_V }{(2\,{\rm Im}\,\tau)^{3}_{}|\eta(\tau)|^{12}_{}}\left[\frac{(2\,{\rm Im}\,\tau)^2_{}}{3}\left|\I H^\prime(\tau) + \frac{3}{2\pi}\widehat{G}_2(\tau,\overline{\tau})H(\tau)\right|^2_{} + \left(\calA -3 \right)|H(\tau)|^2\right] \; ,
    \label{eq:single-ponten-surga}
\end{equation}
which recovers the scalar potential used for modulus stabilisation in Refs.~\cite{Leedom:2022zdm,King:2023snq}.

Next, let us derive the bias term, by considering the dilaton term $\calA$, and relaxing the global supersymmetry limit, i.e., by allowing $\fm$ to be non-zero but still a small parameter. Indeed, there are several terms in the scalar potential that depend on $\calA$, as shown in Eq.~(\ref{eq:single-ponten}), whereas we only turn on the term $\fm \calA |H(\tau)|^2$. The reasons for this selective treatment are as follows. First, the inclusion of $(2\,{\rm Im}\,\tau)^{3\fm}_{}|\eta(\tau)|^{12\fm}_{}$ and $3\fm\,\widehat{G}_2(\tau,\overline{\tau})H(\tau)/(2\pi)$ with a small $\fm$ does not break the degeneracy between two vacua at $\tau = \I$ and $\omega$, which remain as Minkowski vacua. In this regard, they make very small modifications to the potential, given that they are suppressed by $\plk$. Hence we can safely neglect them. Second, the term $-3\fm|H(\tau)|^2$ can break the vacua degeneracy, shifting $\tau = \I$ down to an anti-de Sitter (AdS) vacuum. However, this term  could also induce a non-vanishing gravitational mass, compensating for the difference in the values of the potential at $\tau = \I$ and $\omega$. As a result, the total energy, defined as a sum of the vacuum and gravitational energies, is the same at $\tau = \I$ and $\omega$, ensuring the stability of DWs~\cite{Cvetic:1992bf}. Therefore, the term  $-3|H(\tau)|^2$ can not act as a bias term. Conversely, as mentioned above, the term $\fm \calA |H(\tau)|^2$ comes from the $F$-term of the dilaton, rather than from gravitational corrections. Finally, one can choose relatively large values for $\calA$, e.g., $\calA \gtrsim {\cal O}(10)$, to ensure that the term $\fm \calA |H(\tau)|^2$ dominates over other $\fm$-suppressed terms. As discussed in Sec.~\ref{sec:gw}, such values of $\calA$ are indeed favoured in our model.
Based on the above arguments, we incorporate only the term $\fm \calA |H(\tau)|^2$ into the global supersymmetry potential given in Eq.~(\ref{eq:global-ponten}), namely,
\begin{equation}
\begin{split}
     V(\tau, \overline{\tau}) = & \Lambda^4_V\left[\frac{(2\,{\rm Im}\,\tau)^2_{} }{ 3 }\left|H^\prime(\tau) \right|^2_{} + \fm \calA |H(\tau)|^2\right] \\
     = &\frac{\Lambda^6_W}{\Lambda^2_K}\frac{(2\,{\rm Im}\,\tau)^2_{}}{3}\left| H^\prime(\tau) \right|^2  + \frac{\Lambda^6_W}{\plk^2}\calA|H(\tau)|^2\; ,
\end{split}
\label{eq:ponten-bias-add}
\end{equation}
where $\Lambda^4_V \equiv \Lambda^6_W/\Lambda^2_K$ and $\fm \equiv \Lambda^2_K/\plk^2$ have been used. As $H(\tau)$ takes different values at $\tau = \I$ and $\omega$, the last term in Eq.~(\ref{eq:ponten-bias-add}) can generate a bias term as shown in Eq.~(\ref{eq:Vbias}).

\section{Complete field equations in ${\cal N } = 1$ supergravity} \label{sec:appC}
In this appendix, we present the equations for the modulus field in the presence of a planar thin DW within supergravity theory, and investigate under which conditions they reduce to the global supersymmetry case.

We have mentioned that the DW can result in the dramatic change of the spacetime metric. Assuming the spatial components of the metric parallel to the wall are homogeneous and isotropic in the co-moving frame, we define ${\rm d}s^2 = \alpha(z)({\rm d}t^2-{\rm d}z^2)+\beta(z)(-{\rm d}x^2-{\rm d}y^2)$, with $z$ being the coordinate transverse to the wall. Note that here we are looking for the time-independent metric solution, hence the dependence on time of the metric has been omitted. The first-order field equations can be derived by allowing the variations of corresponding spinors within the superfields to be vanishing. Here we directly give the complete set of equations that the modulus field and the metric should satisfy as follows~\cite{Cvetic:1992bf,Cvetic:1996vr}
\begin{equation}
    \begin{split}
        \frac{\partial \tau}{\partial z}  & = - \zeta \sqrt{\alpha} e^{\calK/(2\plk^2)}\calK^{\tau\overline{\tau}}\frac{|\calW|}{\plk}\frac{D_{\overline{\tau}} \overline{\calW}}{\overline{\calW}} \; , \\
        \frac{\partial \ln \alpha}{\partial z} & = \frac{\partial \ln \beta}{\partial z} = 2\zeta\sqrt{\alpha} e^{\calK/(2\plk^2)} \frac{|\calW|}{\plk^3} \; , \\
        {\rm Im}\left(\frac{\partial \tau}{\partial z}\frac{D_\tau \calW}{\calW} \right) & = 0  \; ,
    \end{split}
    \label{eq:firsteof}
\end{equation}
 with $\zeta = \pm 1$. The second equation of Eq.~\eqref{eq:firsteof} comes from $\partial_z \ln \alpha  = \partial_z \ln \beta = 2 \I e^{-\I \Theta} \sqrt{\alpha} \times e^{\calK/(2\plk^2)} \calW/\plk^3$. Since the metric functions $\alpha$ and $\beta$ are real, the phase $\Theta(z)$ should be $z$-dependent and required to meet $\calW(z) = -\I \zeta e^{{\rm i}\Theta(z)}|\calW(z)|$~\cite{Cvetic:1992bf}. $\zeta$ flips its sign when $\calW(z)$ crosses zero, compensating for a discontinuous shift of $\pi$ in the phase of ${\cal W}$, thereby preserving the continuity of the solution. The last equation of Eq.~(\ref{eq:firsteof}) refers to the geodesic path between two vacua in the supergravity potential space. Keeping $\calK^{\tau \overline{\tau}}_{} = (2\,\imt)^2_{}/(3\Lambda^2_K)$ in mind, we observe from the first two equations of Eq.~(\ref{eq:firsteof}) that the variation of $\ln \alpha$ with $z$ relative to that of $\tau$ is suppressed by a factor $\Lambda_K^2/\plk^2$. In this sense, the differential equation for the metric components is decouple in the global supersymmetry limit $\plk \to \infty$, allowing us to consider the case with a constant spacetime metric.

\bibliographystyle{JHEP}
\bibliography{Ref}

\providecommand{\href}[2]{#2}\begingroup\raggedright\begin{thebibliography}{100}

\bibitem{Xing:2020ijf}
Z.-z.~Xing, \emph{{Flavor structures of charged fermions and massive
  neutrinos}}, \href{https://doi.org/10.1016/j.physrep.2020.02.001}{\emph{Phys.
  Rept.} {\bfseries 854} (2020) 1}
  [\href{https://arxiv.org/abs/1909.09610}{{\ttfamily 1909.09610}}].

\bibitem{King:2013eh}
S.F.~King and C.~Luhn, \emph{{Neutrino Mass and Mixing with Discrete
  Symmetry}}, \href{https://doi.org/10.1088/0034-4885/76/5/056201}{\emph{Rept.
  Prog. Phys.} {\bfseries 76} (2013) 056201}
  [\href{https://arxiv.org/abs/1301.1340}{{\ttfamily 1301.1340}}].

\bibitem{Feruglio:2017spp}
F.~Feruglio, \emph{{Are neutrino masses modular forms?}},  in \emph{{From My
  Vast Repertoire ...}: {Guido Altarelli's Legacy}}, A.~Levy, S.~Forte and
  G.~Ridolfi, eds., pp.~227--266 (2019),
  \href{https://doi.org/10.1142/9789813238053_0012}{DOI}
  [\href{https://arxiv.org/abs/1706.08749}{{\ttfamily 1706.08749}}].

\bibitem{Lauer:1989ax}
J.~Lauer, J.~Mas and H.P.~Nilles, \emph{{Duality and the Role of
  Nonperturbative Effects on the World Sheet}},
  \href{https://doi.org/10.1016/0370-2693(89)91190-8}{\emph{Phys. Lett. B}
  {\bfseries 226} (1989) 251}.

\bibitem{Ferrara:1989bc}
S.~Ferrara, D.~Lust, A.D.~Shapere and S.~Theisen, \emph{{Modular Invariance in
  Supersymmetric Field Theories}},
  \href{https://doi.org/10.1016/0370-2693(89)90583-2}{\emph{Phys. Lett. B}
  {\bfseries 225} (1989) 363}.

\bibitem{Ferrara:1989qb}
S.~Ferrara, .D.~Lust and S.~Theisen, \emph{{Target Space Modular Invariance and
  Low-Energy Couplings in Orbifold Compactifications}},
  \href{https://doi.org/10.1016/0370-2693(89)90631-X}{\emph{Phys. Lett. B}
  {\bfseries 233} (1989) 147}.

\bibitem{Kobayashi:2018vbk}
T.~Kobayashi, K.~Tanaka and T.H.~Tatsuishi, \emph{{Neutrino mixing from finite
  modular groups}},
  \href{https://doi.org/10.1103/PhysRevD.98.016004}{\emph{Phys. Rev. D}
  {\bfseries 98} (2018) 016004}
  [\href{https://arxiv.org/abs/1803.10391}{{\ttfamily 1803.10391}}].

\bibitem{Criado:2018thu}
J.C.~Criado and F.~Feruglio, \emph{{Modular Invariance Faces Precision Neutrino
  Data}}, \href{https://doi.org/10.21468/SciPostPhys.5.5.042}{\emph{SciPost
  Phys.} {\bfseries 5} (2018) 042}
  [\href{https://arxiv.org/abs/1807.01125}{{\ttfamily 1807.01125}}].

\bibitem{Kobayashi:2018scp}
T.~Kobayashi, N.~Omoto, Y.~Shimizu, K.~Takagi, M.~Tanimoto and T.H.~Tatsuishi,
  \emph{{Modular A$_{4}$ invariance and neutrino mixing}},
  \href{https://doi.org/10.1007/JHEP11(2018)196}{\emph{JHEP} {\bfseries 11}
  (2018) 196} [\href{https://arxiv.org/abs/1808.03012}{{\ttfamily
  1808.03012}}].

\bibitem{deAnda:2018ecu}
F.J.~de~Anda, S.F.~King and E.~Perdomo, \emph{{$SU(5)$ grand unified theory
  with $A_4$ modular symmetry}},
  \href{https://doi.org/10.1103/PhysRevD.101.015028}{\emph{Phys. Rev. D}
  {\bfseries 101} (2020) 015028}
  [\href{https://arxiv.org/abs/1812.05620}{{\ttfamily 1812.05620}}].

\bibitem{Okada:2019uoy}
H.~Okada and M.~Tanimoto, \emph{{Towards unification of quark and lepton
  flavors in $A_4$ modular invariance}},
  \href{https://doi.org/10.1140/epjc/s10052-021-08845-y}{\emph{Eur. Phys. J. C}
  {\bfseries 81} (2021) 52} [\href{https://arxiv.org/abs/1905.13421}{{\ttfamily
  1905.13421}}].

\bibitem{Ding:2019zxk}
G.-J.~Ding, S.F.~King and X.-G.~Liu, \emph{{Modular A$_{4}$ symmetry models of
  neutrinos and charged leptons}},
  \href{https://doi.org/10.1007/JHEP09(2019)074}{\emph{JHEP} {\bfseries 09}
  (2019) 074} [\href{https://arxiv.org/abs/1907.11714}{{\ttfamily
  1907.11714}}].

\bibitem{Zhang:2019ngf}
D.~Zhang, \emph{{A modular $A_4$ symmetry realization of two-zero textures of
  the Majorana neutrino mass matrix}},
  \href{https://doi.org/10.1016/j.nuclphysb.2020.114935}{\emph{Nucl. Phys. B}
  {\bfseries 952} (2020) 114935}
  [\href{https://arxiv.org/abs/1910.07869}{{\ttfamily 1910.07869}}].

\bibitem{Kobayashi:2019gtp}
T.~Kobayashi, T.~Nomura and T.~Shimomura, \emph{{Type II seesaw models with
  modular $A_4$ symmetry}},
  \href{https://doi.org/10.1103/PhysRevD.102.035019}{\emph{Phys. Rev. D}
  {\bfseries 102} (2020) 035019}
  [\href{https://arxiv.org/abs/1912.00637}{{\ttfamily 1912.00637}}].

\bibitem{Wang:2019xbo}
X.~Wang, \emph{{Lepton flavor mixing and CP violation in the minimal
  type-(I+II) seesaw model with a modular $A_4$ symmetry}},
  \href{https://doi.org/10.1016/j.nuclphysb.2020.115105}{\emph{Nucl. Phys. B}
  {\bfseries 957} (2020) 115105}
  [\href{https://arxiv.org/abs/1912.13284}{{\ttfamily 1912.13284}}].

\bibitem{Okada:2020rjb}
H.~Okada and M.~Tanimoto, \emph{{Quark and lepton flavors with common modulus
  \ensuremath{\tau} in A4 modular symmetry}},
  \href{https://doi.org/10.1016/j.dark.2023.101204}{\emph{Phys. Dark Univ.}
  {\bfseries 40} (2023) 101204}
  [\href{https://arxiv.org/abs/2005.00775}{{\ttfamily 2005.00775}}].

\bibitem{Yao:2020qyy}
C.-Y.~Yao, J.-N.~Lu and G.-J.~Ding, \emph{{Modular Invariant $A_{4}$ Models for
  Quarks and Leptons with Generalized CP Symmetry}},
  \href{https://doi.org/10.1007/JHEP05(2021)102}{\emph{JHEP} {\bfseries 05}
  (2021) 102} [\href{https://arxiv.org/abs/2012.13390}{{\ttfamily
  2012.13390}}].

\bibitem{Chen:2021zty}
P.~Chen, G.-J.~Ding and S.F.~King, \emph{{SU(5) GUTs with A$_{4}$ modular
  symmetry}}, \href{https://doi.org/10.1007/JHEP04(2021)239}{\emph{JHEP}
  {\bfseries 04} (2021) 239}
  [\href{https://arxiv.org/abs/2101.12724}{{\ttfamily 2101.12724}}].

\bibitem{Kobayashi:2021pav}
T.~Kobayashi, H.~Otsuka, M.~Tanimoto and K.~Yamamoto, \emph{{Modular symmetry
  in the SMEFT}},
  \href{https://doi.org/10.1103/PhysRevD.105.055022}{\emph{Phys. Rev. D}
  {\bfseries 105} (2022) 055022}
  [\href{https://arxiv.org/abs/2112.00493}{{\ttfamily 2112.00493}}].

\bibitem{Kang:2022psa}
D.W.~Kang, J.~Kim, T.~Nomura and H.~Okada, \emph{{Natural mass hierarchy among
  three heavy Majorana neutrinos for resonant leptogenesis under modular
  A$_{4}$ symmetry}},
  \href{https://doi.org/10.1007/JHEP07(2022)050}{\emph{JHEP} {\bfseries 07}
  (2022) 050} [\href{https://arxiv.org/abs/2205.08269}{{\ttfamily
  2205.08269}}].

\bibitem{CentellesChulia:2023osj}
S.~Centelles~Chuli\'a, R.~Kumar, O.~Popov and R.~Srivastava, \emph{{Neutrino
  mass sum rules from modular A4 symmetry}},
  \href{https://doi.org/10.1103/PhysRevD.109.035016}{\emph{Phys. Rev. D}
  {\bfseries 109} (2024) 035016}
  [\href{https://arxiv.org/abs/2308.08981}{{\ttfamily 2308.08981}}].

\bibitem{Penedo:2018nmg}
J.T.~Penedo and S.T.~Petcov, \emph{{Lepton Masses and Mixing from Modular $S_4$
  Symmetry}},
  \href{https://doi.org/10.1016/j.nuclphysb.2018.12.016}{\emph{Nucl. Phys. B}
  {\bfseries 939} (2019) 292}
  [\href{https://arxiv.org/abs/1806.11040}{{\ttfamily 1806.11040}}].

\bibitem{Novichkov:2018ovf}
P.P.~Novichkov, J.T.~Penedo, S.T.~Petcov and A.V.~Titov, \emph{{Modular S$_{4}$
  models of lepton masses and mixing}},
  \href{https://doi.org/10.1007/JHEP04(2019)005}{\emph{JHEP} {\bfseries 04}
  (2019) 005} [\href{https://arxiv.org/abs/1811.04933}{{\ttfamily
  1811.04933}}].

\bibitem{Kobayashi:2019mna}
T.~Kobayashi, Y.~Shimizu, K.~Takagi, M.~Tanimoto and T.H.~Tatsuishi, \emph{{New
  $A_4$ lepton flavor model from $S_4$ modular symmetry}},
  \href{https://doi.org/10.1007/JHEP02(2020)097}{\emph{JHEP} {\bfseries 02}
  (2020) 097} [\href{https://arxiv.org/abs/1907.09141}{{\ttfamily
  1907.09141}}].

\bibitem{Wang:2019ovr}
X.~Wang and S.~Zhou, \emph{{The minimal seesaw model with a modular S$_{4}$
  symmetry}}, \href{https://doi.org/10.1007/JHEP05(2020)017}{\emph{JHEP}
  {\bfseries 05} (2020) 017}
  [\href{https://arxiv.org/abs/1910.09473}{{\ttfamily 1910.09473}}].

\bibitem{Zhang:2021olk}
X.~Zhang and S.~Zhou, \emph{{Inverse seesaw model with a modular S 4 symmetry:
  lepton flavor mixing and warm dark~matter}},
  \href{https://doi.org/10.1088/1475-7516/2021/09/043}{\emph{JCAP} {\bfseries
  09} (2021) 043} [\href{https://arxiv.org/abs/2106.03433}{{\ttfamily
  2106.03433}}].

\bibitem{Novichkov:2018nkm}
P.P.~Novichkov, J.T.~Penedo, S.T.~Petcov and A.V.~Titov, \emph{{Modular A$_{5}$
  symmetry for flavour model building}},
  \href{https://doi.org/10.1007/JHEP04(2019)174}{\emph{JHEP} {\bfseries 04}
  (2019) 174} [\href{https://arxiv.org/abs/1812.02158}{{\ttfamily
  1812.02158}}].

\bibitem{Ding:2019xna}
G.-J.~Ding, S.F.~King and X.-G.~Liu, \emph{{Neutrino mass and mixing with $A_5$
  modular symmetry}},
  \href{https://doi.org/10.1103/PhysRevD.100.115005}{\emph{Phys. Rev. D}
  {\bfseries 100} (2019) 115005}
  [\href{https://arxiv.org/abs/1903.12588}{{\ttfamily 1903.12588}}].

\bibitem{Criado:2019tzk}
J.C.~Criado, F.~Feruglio and S.J.D.~King, \emph{{Modular Invariant Models of
  Lepton Masses at Levels 4 and 5}},
  \href{https://doi.org/10.1007/JHEP02(2020)001}{\emph{JHEP} {\bfseries 02}
  (2020) 001} [\href{https://arxiv.org/abs/1908.11867}{{\ttfamily
  1908.11867}}].

\bibitem{Kobayashi:2023zzc}
T.~Kobayashi and M.~Tanimoto, \emph{{Modular flavor symmetric models}},  7,
  2023 [\href{https://arxiv.org/abs/2307.03384}{{\ttfamily 2307.03384}}].

\bibitem{Ding:2023htn}
G.-J.~Ding and S.F.~King, \emph{{Neutrino mass and mixing with modular
  symmetry}}, \href{https://doi.org/10.1088/1361-6633/ad52a3}{\emph{Rept. Prog.
  Phys.} {\bfseries 87} (2024) 084201}
  [\href{https://arxiv.org/abs/2311.09282}{{\ttfamily 2311.09282}}].

\bibitem{Novichkov:2018yse}
P.P.~Novichkov, S.T.~Petcov and M.~Tanimoto, \emph{{Trimaximal Neutrino Mixing
  from Modular A4 Invariance with Residual Symmetries}},
  \href{https://doi.org/10.1016/j.physletb.2019.04.043}{\emph{Phys. Lett. B}
  {\bfseries 793} (2019) 247}
  [\href{https://arxiv.org/abs/1812.11289}{{\ttfamily 1812.11289}}].

\bibitem{Ding:2019gof}
G.-J.~Ding, S.F.~King, X.-G.~Liu and J.-N.~Lu, \emph{{Modular S$_{4}$ and
  A$_{4}$ symmetries and their fixed points: new predictive examples of lepton
  mixing}}, \href{https://doi.org/10.1007/JHEP12(2019)030}{\emph{JHEP}
  {\bfseries 12} (2019) 030}
  [\href{https://arxiv.org/abs/1910.03460}{{\ttfamily 1910.03460}}].

\bibitem{deMedeirosVarzielas:2020kji}
I.~de~Medeiros~Varzielas, M.~Levy and Y.-L.~Zhou, \emph{{Symmetries and
  stabilisers in modular invariant flavour models}},
  \href{https://doi.org/10.1007/JHEP11(2020)085}{\emph{JHEP} {\bfseries 11}
  (2020) 085} [\href{https://arxiv.org/abs/2008.05329}{{\ttfamily
  2008.05329}}].

\bibitem{Dine:1985rz}
M.~Dine, R.~Rohm, N.~Seiberg and E.~Witten, \emph{{Gluino Condensation in
  Superstring Models}},
  \href{https://doi.org/10.1016/0370-2693(85)91354-1}{\emph{Phys. Lett. B}
  {\bfseries 156} (1985) 55}.

\bibitem{Nilles:1982ik}
H.P.~Nilles, \emph{{Dynamically Broken Supergravity and the Hierarchy
  Problem}}, \href{https://doi.org/10.1016/0370-2693(82)90642-6}{\emph{Phys.
  Lett. B} {\bfseries 115} (1982) 193}.

\bibitem{Ferrara:1982qs}
S.~Ferrara, L.~Girardello and H.P.~Nilles, \emph{{Breakdown of Local
  Supersymmetry Through Gauge Fermion Condensates}},
  \href{https://doi.org/10.1016/0370-2693(83)91325-4}{\emph{Phys. Lett. B}
  {\bfseries 125} (1983) 457}.

\bibitem{Kaplunovsky:1987rp}
V.S.~Kaplunovsky, \emph{{One Loop Threshold Effects in String Unification}},
  \href{https://doi.org/10.1016/0550-3213(88)90526-3}{\emph{Nucl. Phys. B}
  {\bfseries 307} (1988) 145}
  [\href{https://arxiv.org/abs/hep-th/9205068}{{\ttfamily hep-th/9205068}}].

\bibitem{Dixon:1990pc}
L.J.~Dixon, V.~Kaplunovsky and J.~Louis, \emph{{Moduli dependence of string
  loop corrections to gauge coupling constants}},
  \href{https://doi.org/10.1016/0550-3213(91)90490-O}{\emph{Nucl. Phys. B}
  {\bfseries 355} (1991) 649}.

\bibitem{Antoniadis:1991fh}
I.~Antoniadis, K.S.~Narain and T.R.~Taylor, \emph{{Higher genus string
  corrections to gauge couplings}},
  \href{https://doi.org/10.1016/0370-2693(91)90521-Q}{\emph{Phys. Lett. B}
  {\bfseries 267} (1991) 37}.

\bibitem{Antoniadis:1992rq}
I.~Antoniadis, E.~Gava and K.S.~Narain, \emph{{Moduli corrections to gauge and
  gravitational couplings in four-dimensional superstrings}},
  \href{https://doi.org/10.1016/0550-3213(92)90672-X}{\emph{Nucl. Phys. B}
  {\bfseries 383} (1992) 93}
  [\href{https://arxiv.org/abs/hep-th/9204030}{{\ttfamily hep-th/9204030}}].

\bibitem{Antoniadis:1992sa}
I.~Antoniadis, E.~Gava and K.S.~Narain, \emph{{Moduli corrections to
  gravitational couplings from string loops}},
  \href{https://doi.org/10.1016/0370-2693(92)90009-S}{\emph{Phys. Lett. B}
  {\bfseries 283} (1992) 209}
  [\href{https://arxiv.org/abs/hep-th/9203071}{{\ttfamily hep-th/9203071}}].

\bibitem{Kaplunovsky:1995jw}
V.~Kaplunovsky and J.~Louis, \emph{{On Gauge couplings in string theory}},
  \href{https://doi.org/10.1016/0550-3213(95)00172-O}{\emph{Nucl. Phys. B}
  {\bfseries 444} (1995) 191}
  [\href{https://arxiv.org/abs/hep-th/9502077}{{\ttfamily hep-th/9502077}}].

\bibitem{Font:1990nt}
A.~Font, L.E.~Ibanez, D.~Lust and F.~Quevedo, \emph{{Supersymmetry Breaking
  From Duality Invariant Gaugino Condensation}},
  \href{https://doi.org/10.1016/0370-2693(90)90665-S}{\emph{Phys. Lett. B}
  {\bfseries 245} (1990) 401}.

\bibitem{Cvetic:1991qm}
M.~Cvetic, A.~Font, L.E.~Ibanez, D.~Lust and F.~Quevedo, \emph{{Target space
  duality, supersymmetry breaking and the stability of classical string
  vacua}}, \href{https://doi.org/10.1016/0550-3213(91)90622-5}{\emph{Nucl.
  Phys. B} {\bfseries 361} (1991) 194}.

\bibitem{Cicoli:2013rwa}
M.~Cicoli, S.~de~Alwis and A.~Westphal, \emph{{Heterotic Moduli
  Stabilisation}}, \href{https://doi.org/10.1007/JHEP10(2013)199}{\emph{JHEP}
  {\bfseries 10} (2013) 199} [\href{https://arxiv.org/abs/1304.1809}{{\ttfamily
  1304.1809}}].

\bibitem{Gonzalo:2018guu}
E.~Gonzalo, L.E.~Ib\'a\~nez and A.M.~Uranga, \emph{{Modular symmetries and the
  swampland conjectures}},
  \href{https://doi.org/10.1007/JHEP05(2019)105}{\emph{JHEP} {\bfseries 05}
  (2019) 105} [\href{https://arxiv.org/abs/1812.06520}{{\ttfamily
  1812.06520}}].

\bibitem{Kobayashi:2019xvz}
T.~Kobayashi, Y.~Shimizu, K.~Takagi, M.~Tanimoto and T.H.~Tatsuishi,
  \emph{{$A_4$ lepton flavor model and modulus stabilization from $S_4$ modular
  symmetry}}, \href{https://doi.org/10.1103/PhysRevD.100.115045}{\emph{Phys.
  Rev. D} {\bfseries 100} (2019) 115045}
  [\href{https://arxiv.org/abs/1909.05139}{{\ttfamily 1909.05139}}].

\bibitem{Ishiguro:2020tmo}
K.~Ishiguro, T.~Kobayashi and H.~Otsuka, \emph{{Landscape of Modular Symmetric
  Flavor Models}}, \href{https://doi.org/10.1007/JHEP03(2021)161}{\emph{JHEP}
  {\bfseries 03} (2021) 161}
  [\href{https://arxiv.org/abs/2011.09154}{{\ttfamily 2011.09154}}].

\bibitem{Novichkov:2022wvg}
P.P.~Novichkov, J.T.~Penedo and S.T.~Petcov, \emph{{Modular flavour symmetries
  and modulus stabilisation}},
  \href{https://doi.org/10.1007/JHEP03(2022)149}{\emph{JHEP} {\bfseries 03}
  (2022) 149} [\href{https://arxiv.org/abs/2201.02020}{{\ttfamily
  2201.02020}}].

\bibitem{Funakoshi:2024yxg}
S.~Funakoshi, J.~Kawamura, T.~Kobayashi, K.~Nasu and H.~Otsuka, \emph{{Moduli
  stabilization and light axion by Siegel modular forms}},
  \href{https://arxiv.org/abs/2409.19261}{{\ttfamily 2409.19261}}.

\bibitem{Ishiguro:2022pde}
K.~Ishiguro, H.~Okada and H.~Otsuka, \emph{{Residual flavor symmetry breaking
  in the landscape of modular flavor models}},
  \href{https://doi.org/10.1007/JHEP09(2022)072}{\emph{JHEP} {\bfseries 09}
  (2022) 072} [\href{https://arxiv.org/abs/2206.04313}{{\ttfamily
  2206.04313}}].

\bibitem{Leedom:2022zdm}
J.M.~Leedom, N.~Righi and A.~Westphal, \emph{{Heterotic de Sitter beyond
  modular symmetry}},
  \href{https://doi.org/10.1007/JHEP02(2023)209}{\emph{JHEP} {\bfseries 02}
  (2023) 209} [\href{https://arxiv.org/abs/2212.03876}{{\ttfamily
  2212.03876}}].

\bibitem{Knapp-Perez:2023nty}
V.~Knapp-Perez, X.-G.~Liu, H.P.~Nilles, S.~Ramos-Sanchez and M.~Ratz,
  \emph{{Matter matters in moduli fixing and modular flavor symmetries}},
  \href{https://arxiv.org/abs/2304.14437}{{\ttfamily 2304.14437}}.

\bibitem{King:2023snq}
S.F.~King and X.~Wang, \emph{{Modulus stabilization in the multiple-modulus
  framework}}, \href{https://doi.org/10.1103/PhysRevD.110.076026}{\emph{Phys.
  Rev. D} {\bfseries 110} (2024) 076026}
  [\href{https://arxiv.org/abs/2310.10369}{{\ttfamily 2310.10369}}].

\bibitem{Shenker:1990}
S.~Shenker, \emph{{The Strength of Nonperturbative Effects in String Theory}},
  {\emph{Random Surfaces and Quantum Gravity,} (1990) 191}.

\bibitem{Ding:2024neh}
G.-J.~Ding, S.-Y.~Jiang and W.~Zhao, \emph{{Modular invariant slow roll
  inflation}}, \href{https://doi.org/10.1088/1475-7516/2024/10/016}{\emph{JCAP}
  {\bfseries 10} (2024) 016}
  [\href{https://arxiv.org/abs/2405.06497}{{\ttfamily 2405.06497}}].

\bibitem{King:2024ssx}
S.F.~King and X.~Wang, \emph{{Modular invariant hilltop inflation}},
  \href{https://doi.org/10.1088/1475-7516/2024/07/073}{\emph{JCAP} {\bfseries
  07} (2024) 073} [\href{https://arxiv.org/abs/2405.08924}{{\ttfamily
  2405.08924}}].

\bibitem{Cvetic:1992bf}
M.~Cvetic, S.~Griffies and S.-J.~Rey, \emph{{Static domain walls in N=1
  supergravity}},
  \href{https://doi.org/10.1016/0550-3213(92)90649-V}{\emph{Nucl. Phys. B}
  {\bfseries 381} (1992) 301}
  [\href{https://arxiv.org/abs/hep-th/9201007}{{\ttfamily hep-th/9201007}}].

\bibitem{Cvetic:1992cv}
M.~Cvetic and R.L.~Davis, \emph{{Cosmological implications of domain walls due
  to duality invariant moduli sector of superstring vacua}},
  \href{https://doi.org/10.1016/0370-2693(92)91327-6}{\emph{Phys. Lett. B}
  {\bfseries 296} (1992) 316}
  [\href{https://arxiv.org/abs/hep-th/9205060}{{\ttfamily hep-th/9205060}}].

\bibitem{Cvetic:1996vr}
M.~Cvetic and H.H.~Soleng, \emph{{Supergravity domain walls}},
  \href{https://doi.org/10.1016/S0370-1573(96)00035-X}{\emph{Phys. Rept.}
  {\bfseries 282} (1997) 159}
  [\href{https://arxiv.org/abs/hep-th/9604090}{{\ttfamily hep-th/9604090}}].

\bibitem{Cvetic:1991vp}
M.~Cvetic, F.~Quevedo and S.-J.~Rey, \emph{{Stringy domain walls and target
  space modular invariance}},
  \href{https://doi.org/10.1103/PhysRevLett.67.1836}{\emph{Phys. Rev. Lett.}
  {\bfseries 67} (1991) 1836}.

\bibitem{Zeldovich:1974uw}
Y.B.~Zeldovich, I.Y.~Kobzarev and L.B.~Okun, \emph{{Cosmological Consequences
  of the Spontaneous Breakdown of Discrete Symmetry}}, {\emph{Zh. Eksp. Teor.
  Fiz.} {\bfseries 67} (1974) 3}.

\bibitem{Kibble:1976sj}
T.W.B.~Kibble, \emph{{Topology of Cosmic Domains and Strings}},
  \href{https://doi.org/10.1088/0305-4470/9/8/029}{\emph{J. Phys. A} {\bfseries
  9} (1976) 1387}.

\bibitem{Vilenkin:1984ib}
A.~Vilenkin, \emph{{Cosmic Strings and Domain Walls}},
  \href{https://doi.org/10.1016/0370-1573(85)90033-X}{\emph{Phys. Rept.}
  {\bfseries 121} (1985) 263}.

\bibitem{Gelmini:2020bqg}
G.B.~Gelmini, S.~Pascoli, E.~Vitagliano and Y.-L.~Zhou, \emph{{Gravitational
  wave signatures from discrete flavor symmetries}},
  \href{https://doi.org/10.1088/1475-7516/2021/02/032}{\emph{JCAP} {\bfseries
  02} (2021) 032} [\href{https://arxiv.org/abs/2009.01903}{{\ttfamily
  2009.01903}}].

\bibitem{Fu:2024jhu}
B.~Fu, S.F.~King, L.~Marsili, S.~Pascoli, J.~Turner and Y.-L.~Zhou,
  \emph{{Non-Abelian Domain Walls and Gravitational Waves}},
  \href{https://arxiv.org/abs/2409.16359}{{\ttfamily 2409.16359}}.

\bibitem{Cremmer:1982en}
E.~Cremmer, S.~Ferrara, L.~Girardello and A.~Van~Proeyen, \emph{{Yang-Mills
  Theories with Local Supersymmetry: Lagrangian, Transformation Laws and
  SuperHiggs Effect}},
  \href{https://doi.org/10.1016/0550-3213(83)90679-X}{\emph{Nucl. Phys. B}
  {\bfseries 212} (1983) 413}.

\bibitem{Kachru:2003aw}
S.~Kachru, R.~Kallosh, A.D.~Linde and S.P.~Trivedi, \emph{{De Sitter vacua in
  string theory}},
  \href{https://doi.org/10.1103/PhysRevD.68.046005}{\emph{Phys. Rev. D}
  {\bfseries 68} (2003) 046005}
  [\href{https://arxiv.org/abs/hep-th/0301240}{{\ttfamily hep-th/0301240}}].

\bibitem{Vilenkin:1981zs}
A.~Vilenkin, \emph{{Gravitational Field of Vacuum Domain Walls and Strings}},
  \href{https://doi.org/10.1103/PhysRevD.23.852}{\emph{Phys. Rev. D} {\bfseries
  23} (1981) 852}.

\bibitem{Gelmini:1988sf}
G.B.~Gelmini, M.~Gleiser and E.W.~Kolb, \emph{{Cosmology of Biased Discrete
  Symmetry Breaking}},
  \href{https://doi.org/10.1103/PhysRevD.39.1558}{\emph{Phys. Rev. D}
  {\bfseries 39} (1989) 1558}.

\bibitem{Larsson:1996sp}
S.E.~Larsson, S.~Sarkar and P.L.~White, \emph{{Evading the cosmological domain
  wall problem}}, \href{https://doi.org/10.1103/PhysRevD.55.5129}{\emph{Phys.
  Rev. D} {\bfseries 55} (1997) 5129}
  [\href{https://arxiv.org/abs/hep-ph/9608319}{{\ttfamily hep-ph/9608319}}].

\bibitem{Hiramatsu:2013qaa}
T.~Hiramatsu, M.~Kawasaki and K.~Saikawa, \emph{{On the estimation of
  gravitational wave spectrum from cosmic domain walls}},
  \href{https://doi.org/10.1088/1475-7516/2014/02/031}{\emph{JCAP} {\bfseries
  02} (2014) 031} [\href{https://arxiv.org/abs/1309.5001}{{\ttfamily
  1309.5001}}].

\bibitem{Hiramatsu:2012sc}
T.~Hiramatsu, M.~Kawasaki, K.~Saikawa and T.~Sekiguchi, \emph{{Axion cosmology
  with long-lived domain walls}},
  \href{https://doi.org/10.1088/1475-7516/2013/01/001}{\emph{JCAP} {\bfseries
  01} (2013) 001} [\href{https://arxiv.org/abs/1207.3166}{{\ttfamily
  1207.3166}}].

\bibitem{Saikawa:2017hiv}
K.~Saikawa, \emph{{A review of gravitational waves from cosmic domain walls}},
  \href{https://doi.org/10.3390/universe3020040}{\emph{Universe} {\bfseries 3}
  (2017) 40} [\href{https://arxiv.org/abs/1703.02576}{{\ttfamily 1703.02576}}].

\bibitem{Roshan:2024qnv}
R.~Roshan and G.~White, \emph{{Using gravitational waves to see the first
  second of the Universe}},  \href{https://arxiv.org/abs/2401.04388}{{\ttfamily
  2401.04388}}.

\bibitem{Kawasaki:2014sqa}
M.~Kawasaki, K.~Saikawa and T.~Sekiguchi, \emph{{Axion dark matter from
  topological defects}},
  \href{https://doi.org/10.1103/PhysRevD.91.065014}{\emph{Phys. Rev. D}
  {\bfseries 91} (2015) 065014}
  [\href{https://arxiv.org/abs/1412.0789}{{\ttfamily 1412.0789}}].

\bibitem{Smith:2006nka}
T.L.~Smith, E.~Pierpaoli and M.~Kamionkowski, \emph{{A new cosmic microwave
  background constraint to primordial gravitational waves}},
  \href{https://doi.org/10.1103/PhysRevLett.97.021301}{\emph{Phys. Rev. Lett.}
  {\bfseries 97} (2006) 021301}
  [\href{https://arxiv.org/abs/astro-ph/0603144}{{\ttfamily
  astro-ph/0603144}}].

\bibitem{Sendra:2012wh}
I.~Sendra and T.L.~Smith, \emph{{Improved limits on short-wavelength
  gravitational waves from the cosmic microwave background}},
  \href{https://doi.org/10.1103/PhysRevD.85.123002}{\emph{Phys. Rev. D}
  {\bfseries 85} (2012) 123002}
  [\href{https://arxiv.org/abs/1203.4232}{{\ttfamily 1203.4232}}].

\bibitem{Pagano:2015hma}
L.~Pagano, L.~Salvati and A.~Melchiorri, \emph{{New constraints on primordial
  gravitational waves from Planck 2015}},
  \href{https://doi.org/10.1016/j.physletb.2016.07.078}{\emph{Phys. Lett. B}
  {\bfseries 760} (2016) 823}
  [\href{https://arxiv.org/abs/1508.02393}{{\ttfamily 1508.02393}}].

\bibitem{Gouttenoire:2019kij}
Y.~Gouttenoire, G.~Servant and P.~Simakachorn, \emph{{Beyond the Standard
  Models with Cosmic Strings}},
  \href{https://doi.org/10.1088/1475-7516/2020/07/032}{\emph{JCAP} {\bfseries
  07} (2020) 032} [\href{https://arxiv.org/abs/1912.02569}{{\ttfamily
  1912.02569}}].

\bibitem{Caprini:2018mtu}
C.~Caprini and D.G.~Figueroa, \emph{{Cosmological Backgrounds of Gravitational
  Waves}}, \href{https://doi.org/10.1088/1361-6382/aac608}{\emph{Class. Quant.
  Grav.} {\bfseries 35} (2018) 163001}
  [\href{https://arxiv.org/abs/1801.04268}{{\ttfamily 1801.04268}}].

\bibitem{Nilles:1990zd}
H.P.~Nilles, \emph{{Gaugino Condensation and Supersymmetry Breakdown}},
  \href{https://doi.org/10.1142/S0217751X90001744}{\emph{Int. J. Mod. Phys. A}
  {\bfseries 5} (1990) 4199}.

\bibitem{Binetruy:1996uv}
P.~Binetruy and E.~Dudas, \emph{{Gaugino condensation and the anomalous U(1)}},
  \href{https://doi.org/10.1016/S0370-2693(96)01305-6}{\emph{Phys. Lett. B}
  {\bfseries 389} (1996) 503}
  [\href{https://arxiv.org/abs/hep-th/9607172}{{\ttfamily hep-th/9607172}}].

\bibitem{Binetruy:1995ta}
P.~Binetruy and M.K.~Gaillard, \emph{{S duality constraints on effective
  potentials for gaugino condensation}},
  \href{https://doi.org/10.1016/0370-2693(95)01242-7}{\emph{Phys. Lett. B}
  {\bfseries 365} (1996) 87}
  [\href{https://arxiv.org/abs/hep-th/9506207}{{\ttfamily hep-th/9506207}}].

\bibitem{Saririan:1996mr}
K.~Saririan, \emph{{Gaugino condensation with S duality and field theoretical
  threshold corrections}},
  \href{https://doi.org/10.1103/PhysRevD.55.4839}{\emph{Phys. Rev. D}
  {\bfseries 55} (1997) 4839}
  [\href{https://arxiv.org/abs/hep-th/9611061}{{\ttfamily hep-th/9611061}}].

\bibitem{Thrane:2013oya}
E.~Thrane and J.D.~Romano, \emph{{Sensitivity curves for searches for
  gravitational-wave backgrounds}},
  \href{https://doi.org/10.1103/PhysRevD.88.124032}{\emph{Phys. Rev. D}
  {\bfseries 88} (2013) 124032}
  [\href{https://arxiv.org/abs/1310.5300}{{\ttfamily 1310.5300}}].

\bibitem{Schmitz:2020syl}
K.~Schmitz, \emph{{New Sensitivity Curves for Gravitational-Wave Signals from
  Cosmological Phase Transitions}},
  \href{https://doi.org/10.1007/JHEP01(2021)097}{\emph{JHEP} {\bfseries 01}
  (2021) 097} [\href{https://arxiv.org/abs/2002.04615}{{\ttfamily
  2002.04615}}].

\bibitem{Breitbach:2018ddu}
M.~Breitbach, J.~Kopp, E.~Madge, T.~Opferkuch and P.~Schwaller, \emph{{Dark,
  Cold, and Noisy: Constraining Secluded Hidden Sectors with Gravitational
  Waves}}, \href{https://doi.org/10.1088/1475-7516/2019/07/007}{\emph{JCAP}
  {\bfseries 07} (2019) 007}
  [\href{https://arxiv.org/abs/1811.11175}{{\ttfamily 1811.11175}}].

\bibitem{Janssen:2014dka}
G.~Janssen et~al., \emph{{Gravitational wave astronomy with the SKA}},
  \href{https://doi.org/10.22323/1.215.0037}{\emph{PoS} {\bfseries AASKA14}
  (2015) 037} [\href{https://arxiv.org/abs/1501.00127}{{\ttfamily
  1501.00127}}].

\bibitem{Sesana:2019vho}
A.~Sesana et~al., \emph{{Unveiling the gravitational universe at $\mu$-Hz
  frequencies}}, \href{https://doi.org/10.1007/s10686-021-09709-9}{\emph{Exper.
  Astron.} {\bfseries 51} (2021) 1333}
  [\href{https://arxiv.org/abs/1908.11391}{{\ttfamily 1908.11391}}].

\bibitem{LISA:2017pwj}
{\scshape LISA} collaboration, \emph{{Laser Interferometer Space Antenna}},
  \href{https://arxiv.org/abs/1702.00786}{{\ttfamily 1702.00786}}.

\bibitem{Ruan:2018tsw}
W.-H.~Ruan, Z.-K.~Guo, R.-G.~Cai and Y.-Z.~Zhang, \emph{{Taiji program:
  Gravitational-wave sources}},
  \href{https://doi.org/10.1142/S0217751X2050075X}{\emph{Int. J. Mod. Phys. A}
  {\bfseries 35} (2020) 2050075}
  [\href{https://arxiv.org/abs/1807.09495}{{\ttfamily 1807.09495}}].

\bibitem{TianQin:2015yph}
{\scshape TianQin} collaboration, \emph{{TianQin: a space-borne gravitational
  wave detector}},
  \href{https://doi.org/10.1088/0264-9381/33/3/035010}{\emph{Class. Quant.
  Grav.} {\bfseries 33} (2016) 035010}
  [\href{https://arxiv.org/abs/1512.02076}{{\ttfamily 1512.02076}}].

\bibitem{Kawamura:2020pcg}
S.~Kawamura et~al., \emph{{Current status of space gravitational wave antenna
  DECIGO and B-DECIGO}},
  \href{https://doi.org/10.1093/ptep/ptab019}{\emph{PTEP} {\bfseries 2021}
  (2021) 05A105} [\href{https://arxiv.org/abs/2006.13545}{{\ttfamily
  2006.13545}}].

\bibitem{Corbin:2005ny}
V.~Corbin and N.J.~Cornish, \emph{{Detecting the cosmic gravitational wave
  background with the big bang observer}},
  \href{https://doi.org/10.1088/0264-9381/23/7/014}{\emph{Class. Quant. Grav.}
  {\bfseries 23} (2006) 2435}
  [\href{https://arxiv.org/abs/gr-qc/0512039}{{\ttfamily gr-qc/0512039}}].

\bibitem{AEDGE:2019nxb}
{\scshape AEDGE} collaboration, \emph{{AEDGE: Atomic Experiment for Dark Matter
  and Gravity Exploration in Space}},
  \href{https://doi.org/10.1140/epjqt/s40507-020-0080-0}{\emph{EPJ Quant.
  Technol.} {\bfseries 7} (2020) 6}
  [\href{https://arxiv.org/abs/1908.00802}{{\ttfamily 1908.00802}}].

\bibitem{Badurina:2019hst}
L.~Badurina et~al., \emph{{AION: An Atom Interferometer Observatory and
  Network}}, \href{https://doi.org/10.1088/1475-7516/2020/05/011}{\emph{JCAP}
  {\bfseries 05} (2020) 011}
  [\href{https://arxiv.org/abs/1911.11755}{{\ttfamily 1911.11755}}].

\bibitem{Punturo:2010zz}
M.~Punturo et~al., \emph{{The Einstein Telescope: A third-generation
  gravitational wave observatory}},
  \href{https://doi.org/10.1088/0264-9381/27/19/194002}{\emph{Class. Quant.
  Grav.} {\bfseries 27} (2010) 194002}.

\bibitem{Reitze:2019iox}
D.~Reitze et~al., \emph{{Cosmic Explorer: The U.S. Contribution to
  Gravitational-Wave Astronomy beyond LIGO}}, {\emph{Bull. Am. Astron. Soc.}
  {\bfseries 51} (2019) 035}
  [\href{https://arxiv.org/abs/1907.04833}{{\ttfamily 1907.04833}}].

\bibitem{LIGOScientific:2014pky}
{\scshape LIGO Scientific} collaboration, \emph{{Advanced LIGO}},
  \href{https://doi.org/10.1088/0264-9381/32/7/074001}{\emph{Class. Quant.
  Grav.} {\bfseries 32} (2015) 074001}
  [\href{https://arxiv.org/abs/1411.4547}{{\ttfamily 1411.4547}}].

\bibitem{NANOGrav:2023gor}
{\scshape NANOGrav} collaboration, \emph{{The NANOGrav 15 yr Data Set: Evidence
  for a Gravitational-wave Background}},
  \href{https://doi.org/10.3847/2041-8213/acdac6}{\emph{Astrophys. J. Lett.}
  {\bfseries 951} (2023) L8}
  [\href{https://arxiv.org/abs/2306.16213}{{\ttfamily 2306.16213}}].

\bibitem{KAGRA:2021kbb}
{\scshape KAGRA, Virgo, LIGO Scientific} collaboration, \emph{{Upper limits on
  the isotropic gravitational-wave background from Advanced LIGO and Advanced
  Virgo\textquoteright{}s third observing run}},
  \href{https://doi.org/10.1103/PhysRevD.104.022004}{\emph{Phys. Rev. D}
  {\bfseries 104} (2021) 022004}
  [\href{https://arxiv.org/abs/2101.12130}{{\ttfamily 2101.12130}}].

\bibitem{Caprini:2019egz}
C.~Caprini et~al., \emph{{Detecting gravitational waves from cosmological phase
  transitions with LISA: an update}},
  \href{https://doi.org/10.1088/1475-7516/2020/03/024}{\emph{JCAP} {\bfseries
  03} (2020) 024} [\href{https://arxiv.org/abs/1910.13125}{{\ttfamily
  1910.13125}}].

\bibitem{NANOGrav:2023hvm}
{\scshape NANOGrav} collaboration, \emph{{The NANOGrav 15 yr Data Set: Search
  for Signals from New Physics}},
  \href{https://doi.org/10.3847/2041-8213/acdc91}{\emph{Astrophys. J. Lett.}
  {\bfseries 951} (2023) L11}
  [\href{https://arxiv.org/abs/2306.16219}{{\ttfamily 2306.16219}}].

\bibitem{VIRGO:2014yos}
{\scshape VIRGO} collaboration, \emph{{Advanced Virgo: a second-generation
  interferometric gravitational wave detector}},
  \href{https://doi.org/10.1088/0264-9381/32/2/024001}{\emph{Class. Quant.
  Grav.} {\bfseries 32} (2015) 024001}
  [\href{https://arxiv.org/abs/1408.3978}{{\ttfamily 1408.3978}}].

\bibitem{KAGRA:2018plz}
{\scshape KAGRA} collaboration, \emph{{KAGRA: 2.5 Generation Interferometric
  Gravitational Wave Detector}},
  \href{https://doi.org/10.1038/s41550-018-0658-y}{\emph{Nature Astron.}
  {\bfseries 3} (2019) 35} [\href{https://arxiv.org/abs/1811.08079}{{\ttfamily
  1811.08079}}].

\bibitem{EPTA:2023fyk}
{\scshape EPTA, InPTA:} collaboration, \emph{{The second data release from the
  European Pulsar Timing Array - III. Search for gravitational wave signals}},
  \href{https://doi.org/10.1051/0004-6361/202346844}{\emph{Astron. Astrophys.}
  {\bfseries 678} (2023) A50}
  [\href{https://arxiv.org/abs/2306.16214}{{\ttfamily 2306.16214}}].

\bibitem{EPTA:2023xxk}
{\scshape EPTA, InPTA} collaboration, \emph{{The second data release from the
  European Pulsar Timing Array - IV. Implications for massive black holes, dark
  matter, and the early Universe}},
  \href{https://doi.org/10.1051/0004-6361/202347433}{\emph{Astron. Astrophys.}
  {\bfseries 685} (2024) A94}
  [\href{https://arxiv.org/abs/2306.16227}{{\ttfamily 2306.16227}}].

\bibitem{Reardon:2023gzh}
D.J.~Reardon et~al., \emph{{Search for an Isotropic Gravitational-wave
  Background with the Parkes Pulsar Timing Array}},
  \href{https://doi.org/10.3847/2041-8213/acdd02}{\emph{Astrophys. J. Lett.}
  {\bfseries 951} (2023) L6}
  [\href{https://arxiv.org/abs/2306.16215}{{\ttfamily 2306.16215}}].

\bibitem{Xu:2023wog}
H.~Xu et~al., \emph{{Searching for the Nano-Hertz Stochastic Gravitational Wave
  Background with the Chinese Pulsar Timing Array Data Release I}},
  \href{https://doi.org/10.1088/1674-4527/acdfa5}{\emph{Res. Astron.
  Astrophys.} {\bfseries 23} (2023) 075024}
  [\href{https://arxiv.org/abs/2306.16216}{{\ttfamily 2306.16216}}].

\bibitem{Maggiore:1999vm}
M.~Maggiore, \emph{{Gravitational wave experiments and early universe
  cosmology}}, \href{https://doi.org/10.1016/S0370-1573(99)00102-7}{\emph{Phys.
  Rept.} {\bfseries 331} (2000) 283}
  [\href{https://arxiv.org/abs/gr-qc/9909001}{{\ttfamily gr-qc/9909001}}].

\bibitem{Allen:1997ad}
B.~Allen and J.D.~Romano, \emph{{Detecting a stochastic background of
  gravitational radiation: Signal processing strategies and sensitivities}},
  \href{https://doi.org/10.1103/PhysRevD.59.102001}{\emph{Phys. Rev. D}
  {\bfseries 59} (1999) 102001}
  [\href{https://arxiv.org/abs/gr-qc/9710117}{{\ttfamily gr-qc/9710117}}].

\bibitem{Stauffer:1978kr}
D.~Stauffer, \emph{{Scaling theory of percolation clusters}},
  \href{https://doi.org/10.1016/0370-1573(79)90060-7}{\emph{Phys. Rept.}
  {\bfseries 54} (1979) 1}.

\end{thebibliography}\endgroup

\end{document}